\documentclass[a4paper,english,10pt,superscriptaddress,reprint,amsmath,amssymb,aps,twocolumn,floatfix,notitlepage,showkeys,prd]{revtex4-2}
\usepackage{graphicx}% Include figure files
\usepackage{bm}% bold math
\usepackage{dcolumn}% Align table columns on decimal point
\usepackage[colorlinks=true,citecolor=darkred,urlcolor=darkred, pdfborder={0 0 0}]{hyperref}% add hypertext capabilities
\usepackage{setspace}
\usepackage{caption}
\usepackage{subcaption} % To get subfigures
\usepackage{slashed}
\usepackage[normalem]{ulem}
\usepackage{float}
\usepackage[usenames,dvipsnames]{xcolor}  %%%%%%% usenames,dvipsnames required to get BrickRed
\usepackage{microtype}
\usepackage{lipsum}
\usepackage{amsmath}
\usepackage{amssymb}
\usepackage{mathrsfs}
\usepackage{placeins}
\usepackage{lettrine}   % For Starting a paragraph with a big letter, https://ctan.mines-albi.fr/macros/latex/contrib/lettrine/doc/lettrine.pdf
\input Eileen.fd        % For use of Eileen Caps Reguler from The LaTeX Font Catalogue, https://tug.org/FontCatalogue/eileencapsregular/
\newcommand*\initfamily{\usefont{U}{Eileen}{xl}{n}}
\definecolor{darkred}{rgb}{0.6,0,0}
\definecolor{linkcolor}{rgb}{0,0,0.5}
\graphicspath{{figs/}}
\usepackage[T1]{fontenc} % if needed
\usepackage[compat=1.1.0]{tikz-feynhand}
\usepackage{tikz-feynman}
\tikzfeynmanset{compat=1.1.0}
\usepackage{feynmp}
\usepackage{tikzsymbols}
\usepackage{array}% For table's line thickness change
\usepackage{pifont} % For special character using commend \ding{}

\definecolor{linkcolor}{rgb}{0,0,0.5}
\usepackage{orcidlink}
\begin{document}
\preprint{2601.xxxx}
%
% \title{\boldmath \color{BrickRed}Neutrino Mixing and Mass Matrix Patterns}
\title{\boldmath \color{BrickRed}Seesaw-I and II Hybrid $T^{\prime}$ Symmetric Neutrino Masses with Mass Selection Rule}
\author{Takaaki Nomura \orcidlink{0000-0002-0864-8333}}
\email{nomura@scu.edu.cn}
\affiliation{College of Physics, \href{https://ror.org/011ashp19}{Sichuan University}, Chengdu 610065, China}
\author{Oleg Popov \orcidlink{0000-0002-0249-8493}}
%\footnotetext{Corresponding author.}
\email{opopo001@ucr.edu (corresponding author)}
\affiliation{Faculty of Physics and Mathematics, \href{https://ror.org/02q963474}{Shenzhen MSU-BIT University}, Shenzhen 518172, China}
\date{\today}% It is always \today, today,
\begin{abstract}
The present manuscript studies a recently proposed new neutrino mixing scheme with a neutrino mass selection rule, $m_{\nu}^{13} + m_{\nu}^{22} + m_{\nu}^{31} = 0$, among a neutrino mass matrix elements. The neutrino mass matrix texture is achieved by means of $T^{\prime}$ flavor discrete symmetry extension of the Standard Model gauge group. The model realizing the neutrino mixing pattern consists of a combination of type-I and II seesaw mechanisms in the minimal possible extension of the Standard Model. Stringent predictions are obtained for normal and inverted neutrino mass orderings. Some important predictions include $\alpha_2 \approx 2\pi$ (best fit), $m_{\text{lightest}} \approx 3 - 4 \times 10^{-4}$~eV, and $m_{ee}^{0\nu\beta\beta} \approx 3 \times 10^{-3}$~eV for normal neutrino mass ordering; $110^\circ < \alpha_1 < 170^\circ$, $40(220)^\circ < \alpha_2 < 60(240)^\circ$, $240^\circ < \delta_{\text{Dirac}} < 310^\circ$, and $m_{\text{lightest}} \approx 2 \times 10^{-5} - 7 \times 10^{-3}$~eV for inverted neutrino mass ordering.
\end{abstract}
%Use showkeys class option if keyword display desired
\maketitle % Comment out is "notitlepage" is included in the document definition.
%\tableofcontents
%--------------------------------------------------------------------------------------------------------------------------------------------------------------------------------------------------------
%
\section{Introduction}
\label{sec:intro}
\lettrine[lines=4,findent=-1.2cm]{\normalfont\initfamily \fontsize{17mm}{10mm}\selectfont U \normalfont\initfamily}{}nderstanding the nature of a neutrino sector is one of the important challenges in particle physics. 
The observations of neutrino oscillations revealed non-zero neutrino masses and lepton mixing that are described by the Pontecorvo-Maki-Nakagawa-Sakata (PMNS) mixing matrix~\cite{Pontecorvo:1957cp,Pontecorvo:1957qd,Maki:1962mu}.
Theoretically, tiny neutrino masses can be generated in various ways, such as seesaw mechanisms~\cite{Yanagida:1979gs, Minkowski:1977sc,Mohapatra:1979ia,Schechter:1980gr,Magg:1980ut,Cheng:1980qt,Mohapatra:1980yp,Foot:1988aq}.

The observed neutrino data could be understood by a structure of neutrino mass matrix. For example, we can fit the data using mass matrices that possess some types of two-zero textures~\cite{Frampton:2002yf,Fritzsch:2011qv,Meloni:2014yea}.  
Recently a new type of structure is proposed to fit neutrino data~\cite{Chakraborty:2024eki} that is 
\begin{align}
    \label{eq:mnu_pattern}
    m_\nu &= \left(
    \begin{matrix}
        A & B & F \\
        B & -2F & G \\
        F & G & J 
    \end{matrix} \right).
\end{align}
The structure satisfies the $2m_\nu^{13} + m_\nu^{22} = 0$ sum rule and provides some predictions for neutrino observables (see also refs.~\cite{Hyodo:2024szv,Chakraborty:2024hhq,Dery:2024lem} for other recent approaches related to neutrino mass matrix pattern).
However, it is quite challenging to realize the structure in terms of a renormalizable model.

In this work, we propose a UV complete model realizing the structure of Eq.~\eqref{eq:mnu_pattern} by combining the type-I and type-II seesaw mechanisms under $T'$ flavor symmetry.
We can achieve the structure by introducing right-handed neutrino and scalar fields assigning appropriate representations under gauge and $T'$ symmetries.  
We then show allowed parameters in the model by fitting neutrino data and predictions for neutrino observables. 

%\vspace{2cm}
%Motivation papers~\cite{Chakraborty:2024eki,Hyodo:2024szv,Chakraborty:2024hhq,Dery:2024lem}.

The paper is organized as follows: Sec.~\ref{sec:model} describes the model and its content, neutrino mass generation is given in Sec.~\ref{sec:m_nu}, constraints on the model's parameters are presented in Sec.~\ref{sec:constraints}, results with discussion are given in Sec.~\ref{sec:results_discussion}, and Sec.~\ref{sec:conclusion} concludes the paper.
%
%-------------------------------------------------------------------------------------------------
%
\section{Model}
\label{sec:model}
The Model is constructed by extending the SM gauge group with $T^\prime$ discrete flavor symmetry~\footnote{Detailed review of $T^{\prime}$ discrete symmetry can be found in~\cite{Ishimori:2010au})}. SM field content is amended with several scalar triplets, $\Delta$, in order to generate tiny neutrino masses in the framework of type-II seesaw mechanism. $\Delta$'s transform as $3,1,$ and $1^{\prime\prime}$ under $T^\prime$ flavor symmetry. At this stage of construction the neutrino mass matrix takes the form of Eq.~\eqref{eq:mnu_pattern} with $A=G$ and $B=J$ degeneracy and $m_{\nu}^{13} + m_{\nu}^{22} + m_{\nu}^{31} = 0$ constraint build in. These all are due to the properties of $T^\prime$ discrete symmetry and assigned fields' representations under it. In order to break the aforementioned degeneracies and obtain Eq.~\eqref{eq:mnu_pattern} in the exact form, we proceed by adding SM singlet sterile neutrinos, $N$, in order to generate tiny neutrino masses in the framework of type-I seesaw mechanism. Furthermore, $H$, SM Higgs generates the quark masses, while $H_d\sim(1,2,1/2,3)$ is needed to obtain three independent charged lepton masses where the numbers in the bracket indicate representations/charge under $SU(3)_c$, $SU(2)_L$, $U(1)_Y$ and $T'$. $H_u\sim(1,2,1/2,2^\prime)$ is included as part of the type-I seesaw mechanism, while two copies of $S$, the only SM singlet scalar in the model, are required to generate the seesaw scale Majorana mass for heavy neutrinos.

The complete field content of the model and particle representations are given in Tab.~\ref{tab:model_fields}. 
\begin{table}[h]
    \centering
    \begin{tabular}{ccccc}
        \hline \hline
        Fields & $SU(3)_c$ & $SU(2)_L$ & $U(1)_Y$ & $T^\prime$ \\ \hline
        $Q$ & $\pmb{3}$ & $\pmb{2}$ & $~\frac{1}{6}$ & $\pmb{1}$ \\
        $\Bar{u}$ & $\pmb{\Bar{3}}$ & $\pmb{1}$ & $-\frac{2}{3}$ & $\pmb{1}$ \\
        $\Bar{d}$ & $\pmb{\Bar{3}}$ & $\pmb{1}$ & $~\frac{1}{3}$ & $\pmb{1}$ \\
        $L$ & $\pmb{1}$ & $\pmb{2}$ & $-\frac{1}{2}$ & $\pmb{3}$ \\
        $\Bar{e}$ & $\pmb{1}$ & $\pmb{1}$ & $1$ & $\pmb{1^\prime}$,$\pmb{1}$,$\pmb{1^{\prime\prime}}$ \\
        $N$ & $\pmb{1}$ & $\pmb{1}$ & $0$ & $\pmb{2}$ \\ \hline
        $H$ & $\pmb{1}$ & $\pmb{2}$ & $~\frac{1}{2}$ & $\pmb{1}$ \\ 
        $H_u$ & $\pmb{1}$ & $\pmb{2}$ & $~\frac{1}{2}$ & $\pmb{2^\prime}$ \\ 
        $H_d$ & $\pmb{1}$ & $\pmb{2}$ & $-\frac{1}{2}$ & $\pmb{3}$ \\ 
        $\Delta_{3,1,1^{\prime\prime}}$ & $\pmb{1}$ & $\pmb{3}$ & $1$ & $\pmb{3}$,$\pmb{1}$,$\pmb{1^{\prime\prime}}$ \\ 
        $S_{1,2}$ & $\pmb{1}$ & $\pmb{1}$ & $0$ & $\pmb{3}$ \\ 
        \hline \hline
    \end{tabular}
    \caption{Model field content.}
    \label{tab:model_fields}
\end{table}

The model Lagrangian is given in Eq.~\eqref{eq:lag}. The sole role of $H$ is to give masses to the SM quarks, $H_d$ sources the charged lepton masses, similar to the $\mathcal{A}_4$ discrete symmetry based models, while $H_u$ and $\Delta$'s generate the neutrino masses by the means of  type-I and type-II seesaw mechanisms. Finally, Majorana masses of the heavy neutrinos are
obtained via the $S_{1,2}$ scalar vacuum expectation values (VEVs).

\begin{subequations}
    \label{eq:lag}
    \begin{align}
        \label{eq:lag_yuk}
        \mathcal{L} &= - Q \Bar{u} H - Q \Bar{d} H^\dagger \\
        &- Y_e \left(H_d L\right)_{\pmb{1^{\prime\prime}}} \Bar{e}_1 - Y_\mu \left(H_d L\right)_{\pmb{1}} \Bar{e}_2 - Y_\tau \left(H_d L\right)_{\pmb{1^{\prime}}} \Bar{e}_3 \nonumber \\
        &- Y_1 \left(L L \right)_{\pmb{3}} \Delta_3 - Y_2 \left(L L \right)_{\pmb{1}} \Delta_1 - Y_3 \left(L L \right)_{\pmb{1^\prime}} \Delta_{1^{\prime\prime}} \nonumber \\
        &- Y_4 L \left(N H_u \right)_{\pmb{3}} - Y_5 \left(N N \right)_{\pmb{3}} S_1  - Y_6 \left(N N \right)_{\pmb{3}} S_2 + \text{h.c.} \nonumber \\
        \label{eq:lag_V}
        V &= \displaystyle\sum_{H,H_{u,d},\Delta_{3,1,1^{\prime\prime}},S_{1,2}} \left[\mu_{X_i,X_j} X_i^\dagger X_j \right. \\
        &\left.+ \lambda_{X_i X_j} \left(X_i^\dagger X_j\right)^2\right] + \mu_{S_{ij}}^2 S_i S_j + \mu_{\Delta_1}^2 \Delta_1 \Delta_1 \nonumber \\
        &+ \mu_{\Delta_3}^2 \Delta_3 \Delta_3 + \mu_{HH1} H \Delta_1^\dagger H + \mu_{HdS} H H_d S_{1,2} \nonumber \\
        &+ \mu_{Hd3} H^\dagger \Delta_3 H_d + \mu_{uu3} H_u \Delta_3^\dagger H_u + \mu_{dd1} H_d \Delta_1 H_d \nonumber \\
        &+ \mu_{dd1^{\prime\prime}} H_d \Delta_{1^{\prime\prime}} H_d + \mu_{dd3} H_d \Delta_3 H_d + \mu_{SSS}^{ijk} S_i S_j S_k \nonumber \\
        &+ \mu_{333} \Delta_3 \Delta_3 \Delta_3 + \mu_{111} \Delta_1 \Delta_1 \Delta_1 + \mu_{1^{\prime\prime}1^{\prime\prime}1^{\prime\prime}} \Delta_{1^{\prime\prime}} \Delta_{1^{\prime\prime}} \Delta_{1^{\prime\prime}} \nonumber \\
        &+ \mu_{331} \Delta_3 \Delta_3 \Delta_1 + \mu_{331^{\prime\prime}} \Delta_3 \Delta_3 \Delta_{1^{\prime\prime}} + \mu_{33S_{1,2}} \Delta_3 \Delta_3 S_{1,2} \nonumber \\
        &+ \dots + \text{h.c.}. \nonumber
    \end{align}
\end{subequations}

In Eq.~\eqref{eq:lag_V} the canonical terms stand for scalar potential terms that consist of multiplications of invariant pairs like ($H^\dagger H$). The indices $i,j$ indicate the generation of SM singlet scalars $S$. We have shown non-canonical terms up to dimension three, while dimension four non-canonical terms, due to a large number of such terms, will be given elsewhere.

In the following section of the manuscript we will study the neutrino mass matrix and how it is obtained from the Lagrangian given in Eq.~\eqref{eq:lag_yuk}.
%
%-------------------------------------------------------------------------------------------------
%
\section{Neutrino Masses}
\label{sec:m_nu}
From the Yukawa Lagrangian in Eq.~\eqref{eq:lag_yuk}, the expanded form of the Lagrangian part that contributes to the neutrino masses can be written as
    \begin{align}
        \label{eq:lag_mnu}
        \mathcal{L}_{\nu} &= 
        - Y_1 \left[ \left(2 L_1 L_1 - L_2 L_3 - L_3 L_2\right) \Delta_3^1 \right. \\
        &+ \left(2 L_2 L_2 - L_1 L_3 - L_3 L_1\right) \Delta_3^2 \nonumber \\
        &\left.+ \left(2 L_3 L_3 - L_1 L_2 - L_2 L_1\right) \Delta_3^3 \right] \nonumber \\
        &- Y_2 \left( L_1 L_1 + L_2 L_3 + L_3 L_2 \right) \Delta_1 \nonumber \\
        &- Y_3 \left( L_3 L_3 + L_1 L_2 + L_2 L_1 \right) \Delta_{1^{\prime\prime}} \nonumber \\
        &- \frac{Y_4}{\sqrt{2}} \left[ \left( \sqrt{2} N_1 L_2 - N_2 L_3 \right) H_u^1 \right. \nonumber \\
        &\left.-\left( \sqrt{2} N_2 L_1 + N_1 L_3 \right) H_u^2 \right] \nonumber \\
        &- Y_5 \left[ 1/\sqrt{2} \left( N_1 N_2 + N_2 N_1 \right) S_1^1 \right. \nonumber \\
        &\left.+ N_2 N_2 S_1^2 - N_1 N_1 S_1^3 \right] \nonumber \\
        &- Y_6 \left[ 1/\sqrt{2} \left( N_1 N_2 + N_2 N_1 \right) S_2^1 \right. \nonumber \\
        &\left.+ N_2 N_2 S_2^2 - N_1 N_1 S_2^3 \right] + \text{h.c.}, \nonumber
    \end{align}
where $T\prime$ multiplets are defined as $L = \left(L_1,L_2,L_3\right)$, $\Delta_3 = (\Delta_3^1,\Delta_3^2,\Delta_3^3)$, $H_u = (H_u^1, H_u^2)$, $S_{1(2)} = (S_{1(2)}^1, S_{1(2)}^2,S_{1(2)}^3)$, and $N = (N_1, N_2)$; the multiplication rules for $T'$ representations can be referred to, for example, ref.~\cite{Ishimori:2010au}.
For realizing the texture of Eq.~\eqref{eq:mnu_pattern}, we assume the configuration of VEVs of scalar fields such that 
\begin{align}
& \langle \Delta_3 \rangle = (0,v_\Delta/\sqrt{2},0), \quad 
\langle H_u \rangle = (0, v_u/\sqrt{2}),  \\
& \langle S_1 \rangle = (0,v_{S_1}/\sqrt{2},0), \quad
\langle S_2 \rangle = (0,0,v_{S_2}/\sqrt{2}), \nonumber
\end{align}
and $\langle \Delta_1 \rangle = v_{\Delta_1}/\sqrt{2}$ and $\langle \Delta_{1''} \rangle = v_{\Delta_{1''}}/\sqrt{2}$.
Then we obtain the mass terms relevant to the active neutrino mass as
\begin{align}
\mathcal{L}_{m_\nu} = & -\sqrt{2}Y_1 v_\Delta \left(2 \overline{\nu_2^c} \nu_2 - \overline{\nu_1^c} \nu_3 - \overline{\nu_3^c} \nu_1 \right) \label{eq:nu-mass-terms}  \\
& - \sqrt{2} Y_2 v_{\Delta_{1'}} \left( \overline{\nu_1^c} \nu_1 + \overline{\nu_2^c} \nu_3 + \overline{\nu_3^c} \nu_2 \right)  \nonumber \\
& - \sqrt{2} Y_3 v_{\Delta_{1''}} \left( \overline{\nu_3^c} \nu_3 + \overline{\nu_1^c} \nu_2 + \overline{\nu_2^c} \nu_1 \right)  \nonumber \\
&+ Y_4 v_u (\sqrt{2} \overline{N_2^c} \nu_2 + \overline{N_1^c} \nu_3 ) \nonumber \\
&- \sqrt2 Y_5 v_{S_1} \overline{N_2^c} N_2 
+\sqrt{2} Y_6 v_{S_2} \overline{N_1^c} N_1, \nonumber 
\end{align}
where $\nu_i$ is left-handed neutrino field.
The last two lines in Eq.~\eqref{eq:nu-mass-terms} generate active Majorana neutrino masses via the usual type-I seesaw mechanism, while the first three lines directly provide the mass terms as in the canonical type-II seesaw scheme.
Thus, these type-I and type-II seesaw contributions give us the texture of Eq.~\eqref{eq:mnu_pattern}.
Therefore, the matrix elements are written as follows
\begin{align}
\label{eq:yuk_abfgj_relation}
& A = -\sqrt{2} Y_2 v_{\Delta_{1'}} + \frac{1}{2\sqrt{2}} \frac{Y_4^2 v_u^2}{Y_5 v_{s_1} - Y_2 v_{\Delta_{1'}}}, \\
& B = -\sqrt{2} Y_3 v_{\Delta_{1''}}, \quad F = \sqrt{2} Y_1 v_{\Delta}, \quad G = -\sqrt{2} Y_2 v_{\Delta_{1'}}, \nonumber \\
& J = -\sqrt{2} Y_3 v_{\Delta_{1''}} - \frac{1}{4\sqrt{2}} \frac{Y_4^2 v_u^2}{Y_6 v_{s_2} + Y_3 v_{\Delta_{1''}}}.
\nonumber
\end{align}
For this neutrino mass matrix structure, we numerically analyze masses, mixing angles, CP-violating phases and effective neutrino mass for neutrinoless double beta decay, taking into account various constraints in the next section.

%
%-------------------------------------------------------------------------------------------------
%
\section{Constraints}
\label{sec:constraints}
Described here are the experimental and theoretical constraints that are relevant for the numerical analysis (see sec.~\ref{sec:results_discussion}) performed for the $T^{\prime}$ discrete symmetry based model considered in sec.~\ref{sec:model}.

The following experimental and theoretical constraints are considered
\begin{itemize}
    \item $m_{\nu}^{13} + m_{\nu}^{22} + m_{\nu}^{31} = 0$~\cite{Chakraborty:2024eki} neutrino masses' selection rule is implemented into the manuscripts model construction.
    \item solar neutrino mass squared splitting is taken as $\Delta m_{21}^2 = \left(7.41\pm 0.21\right)\times 10^{-5}$~eV$^2$~\cite{Esteban:2024eli} for normal and inverted neutrino mass orderings.
    \item atmospheric neutrino mass squared splittings are taken as $\Delta m_{32}^2 = \left(2.437\pm 0.028\right)\times 10^{-3}$~eV$^2$~\cite{Esteban:2024eli} for normal neutrino mass ordering and $\Delta m_{32}^2 = \left(-2.498\pm 0.032\right)\times 10^{-3}$~eV$^2$~\cite{Esteban:2024eli} for inverted neutrino mass ordering.
    \item lepton mixing angle $\theta_{12} = 33.41^\circ \pm 0.75^\circ$~\cite{Esteban:2024eli} for normal and inverted neutrino mass ordering.
    \item lepton mixing angle $\theta_{23} = 49.1^\circ \pm 1.3^\circ$~\cite{Esteban:2024eli} for normal neutrino mass ordering and $\theta_{23} = 49.5^\circ \pm 1.2^\circ$~\cite{Esteban:2024eli} for inverted neutrino mass ordering.
    \item lepton mixing angle $\theta_{13} = 8.54^\circ \pm 0.12^\circ$~\cite{Esteban:2024eli} for normal neutrino mass ordering and $\theta_{13} = 8.57^\circ \pm 0.12^\circ$~\cite{Esteban:2024eli} for inverted neutrino mass ordering.
    \item $\delta_d$ Dirac phase is taken as $\delta_{\text{Dirac}} = 197^{\circ+42}_{~-25}$~\cite{Esteban:2024eli} for normal neutrino mass ordering and $\delta_{\text{Dirac}} = 286^{\circ+27}_{~-32}$~\cite{Esteban:2024eli} for inverted neutrino mass ordering.
    \item neutrino mass sum $\sum m_{\nu} < 0.06$~eV~\cite{Wang:2024hen} cosmological constraint is considered which is a combination of cosmic microwave background (CMB), DESI 2024 Baryon Acoustic Oscillations (BAO), Supernovae Ia, Gamma Ray Bursts (GRB), etc.
    \item neutrino mass sum $\sum m_{\nu} < 0.072 (0.113)$~eV~\cite{DESI:2024mwx,DESI:2024hhd}  for $\sum m_{\nu} > 0 (0.059)$~eV prior from a recent Dark Energy Spectroscopic Instrument Data Release 1 of 2024 (DESI 2024 DR1) cosmological constraint is considered.
    \item lower bounds from neutrino oscillations on the neutrino mass sum are considered $\sum m_{\nu} > 0.059 (0.0992)$~eV~\cite{ParticleDataGroup:2024cfk} for normal (inverted) neutrino mass ordering.
    \item neutrino mass sum $\sum m_{\nu} < 0.265 (0.218)$~eV~\cite{Planck:2018vyg} bounds from PLANCK 2018 CMB Results with(out) CMB lensing are considered.
    \item neutrinoless double beta decay $m_{ee}^{0\nu\beta\beta} < 0.028 - 0.122$~eV~\cite{KamLAND-Zen:2024eml} from KamLAND-Zen Collaboration's latest release of results and $m_{ee}^{0\nu\beta\beta} < 0.010 - 0.020$~eV~\cite{Agostini:2017jim,Giuliani:2019uno} from KamLAND-Zen Collaboration's future sensitivity range.
    \item lower bound from neutrino oscillations on the neutrinoless double beta decay is considered and $m_{ee}^{0\nu\beta\beta} > 0.016$~eV~\cite{ParticleDataGroup:2024cfk} constraint is applied only for inverted neutrino mass ordering.
    \item upper bound on $m_{\nu_{e}}^{\text{eff}} < 0.45$~eV~\cite{KATRIN:2024cdt} from KATRIN Collaboration's 2024 results and KATRIN Experiment's future target sensitivity $m_{\nu_{e}}^{\text{eff}} < 0.2$~eV~\cite{KATRIN:2021uub,Drexlin:2013lha} are considered.
    \item lower bound from neutrino oscillations on the kinematic search for the $\nu_e$ (a.k.a. $m_{\nu_{e}}^{\text{eff}}$) is considered and $m_{\nu_{e}}^{\text{eff}} > 0.0085 (0.0480)$~eV~\cite{ParticleDataGroup:2024cfk} constraint is applied for normal (inverted) neutrino mass ordering.
\end{itemize}

After listing all relevant constraints for model described in Sec.~\ref{sec:model}, the manuscript continues with the results and discussion in the next section of the paper.
%
%-------------------------------------------------------------------------------------------------
%
\section{Results and Discussion}
\label{sec:results_discussion}
This part of the manuscript presents the results of the numerical analysis of the model described in sec.~\ref{sec:model}. All results and generated plots are subject to the constraints given in Sec.~\ref{sec:constraints}.

The color labeling of all plots for normal neutrino mass ordering is as follows: gray points were excluded by one of the experimental constraints, red points satisfy all experimental constraints and fall within 5$\sigma$ intervals of three lepton mixing angles, yellow points satisfy all experimental constraints and fall within 3$\sigma$ intervals of three lepton mixing angles, and finally green points satisfy all experimental constraints and fall within 1$\sigma$ intervals of three lepton mixing angles.

Similarly, but not exactly the same, the color labeling of all plots for inverted neutrino mass ordering is as written next: gray points were excluded by one of the experimental constraints, red points satisfy all experimental constraints and fall within 5$\sigma$ intervals of three lepton mixing angles, yellow points satisfy all experimental constraints and fall within 3$\sigma$ intervals of three lepton mixing angles. Finally, green points satisfy all experimental constraints and fall within 1$\sigma$ intervals of three lepton mixing angles, and lastly dark blue points indicate the region of parameter space and observables' range that is within the KamLAND-Zen Experiment's~\cite{KamLAND-Zen:2024eml} future sensitivity reach.

Every plot is labeled with NO(IO), which stands for Normal (Inverted) neutrino mass ordering, respectively.

Finally, all points shown in any of the plots, of any color, satisfy the neutrino masses' selection rule written as $m_{\nu}^{13} + m_{\nu}^{22} + m_{\nu}^{31} = 0$~\cite{Chakraborty:2024eki}. For those regions, of any plot shown in this section that lacks any points of any aforementioned color, it means that for those regions no solution was found for the $m_{\nu}^{13} + m_{\nu}^{22} + m_{\nu}^{31} = 0$ neutrino masses' selection rule constraint.

\begin{figure*}[h]
    \centering
    \begin{subfigure}[t]{0.45\textwidth}
        \includegraphics[width=0.95\textwidth]{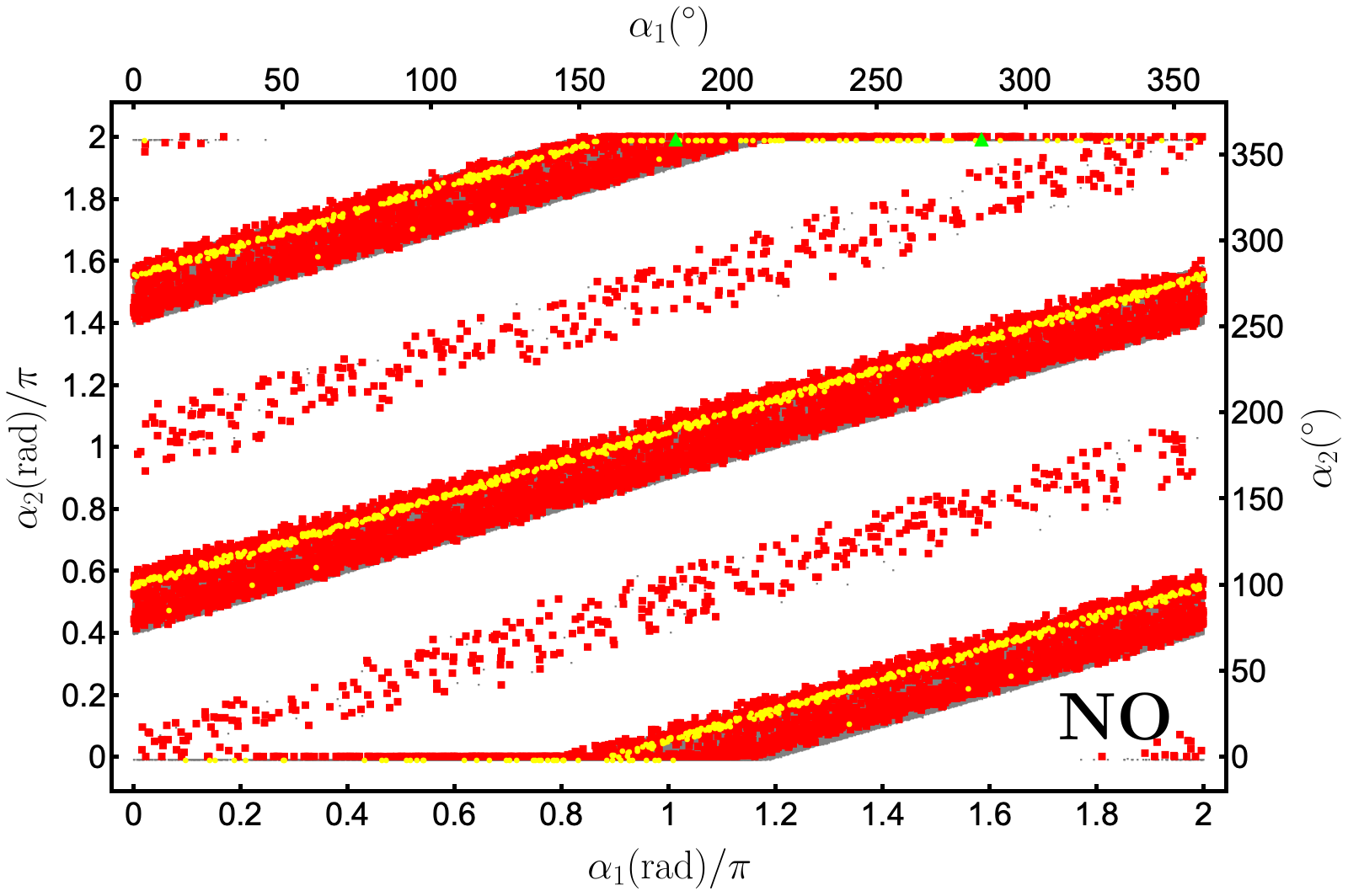}
        \caption{Normal neutrino mass ordering. Gray color indicates those points that are excluded by experimental constraints. Red, yellow, green coloring stands for $5\sigma$, $3\sigma$, and $1\sigma$ confidence intervals with respect to lepton mixing $\theta_{12}$, $\theta_{32}$, and $\theta_{13}$ angles, respectively.}
        \label{fig:alpha1_alpha2_NO}
    \end{subfigure}
    \begin{subfigure}[t]{0.45\textwidth}
        \includegraphics[width=0.95\textwidth]{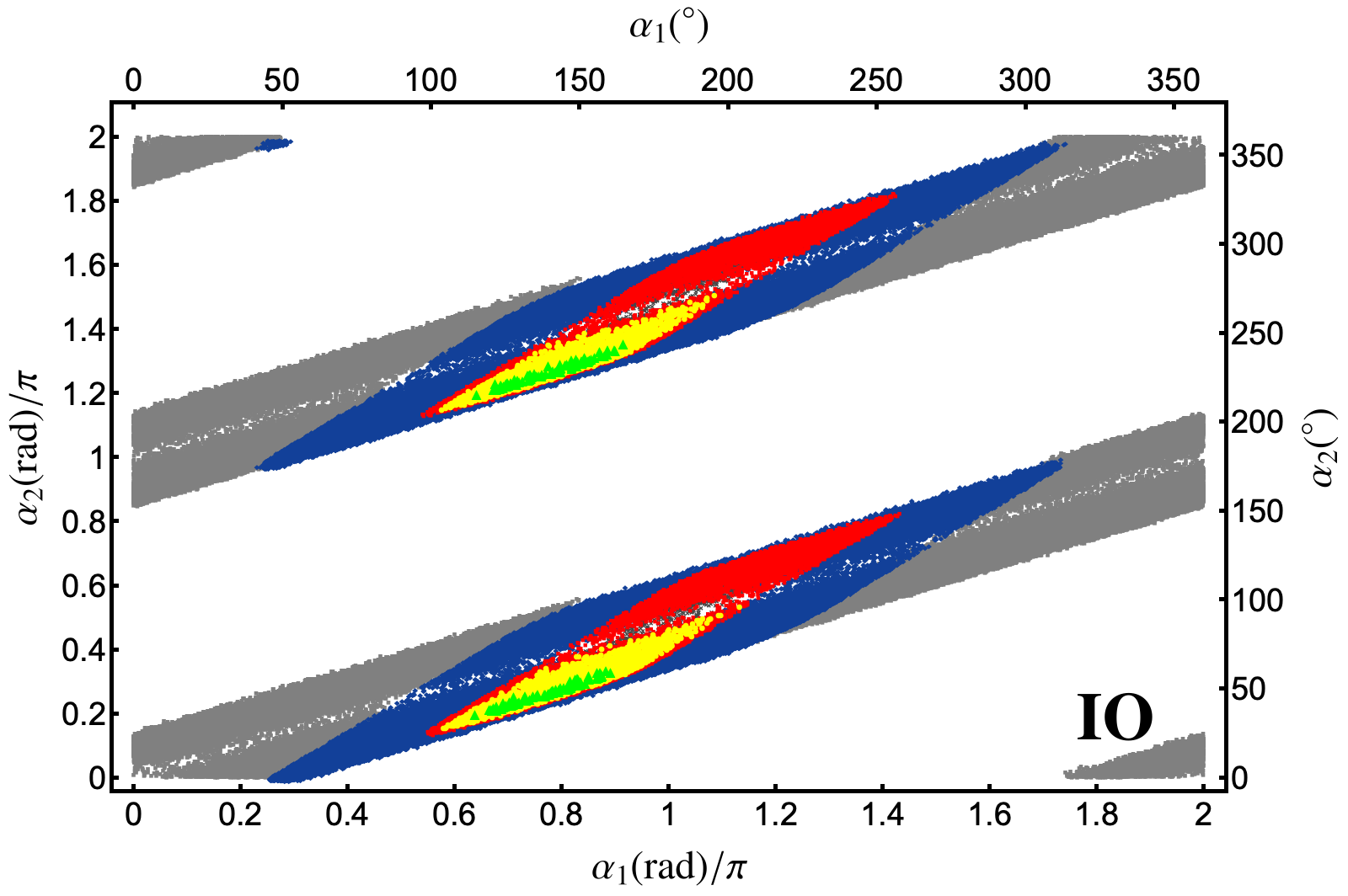}
        \caption{Inverted neutrino mass ordering. Gray color indicates those points that are excluded by experimental constraints. Dark blue colored points are those that will be within future neutrinoless double beta decay sensitivity. Red, yellow, green coloring stands for $5\sigma$, $3\sigma$, and $1\sigma$ confidence intervals with respect to lepton mixing $\theta_{12}$, $\theta_{32}$, and $\theta_{13}$ angles, respectively.}
        \label{fig:alpha1_alpha2_IO}
    \end{subfigure}
    \caption{$\alpha_1$ vs $\alpha_2$ correlation.}
    \label{fig:alpha1_alpha2_plots}
\end{figure*}

%\alpha_1 vs \alpha_2 plots description, Fig. 1
A valid solution and correlation for the $\alpha_1$ and $\alpha_2$ Majorana phases is depicted in Fig.~\ref{fig:alpha1_alpha2_plots}. The $\alpha_1$ vs $\alpha_2$ results for normal neutrino mass ordering are given in Fig.~\ref{fig:alpha1_alpha2_NO}, where a linear correlation is present between the two Majorana phases. With a period of $\pi$ for $\alpha_1$ and a period of $\pi/2$ for $\alpha_2$. A 1 $\sigma$ solutions are localized in the neighborhood $2\pi$ value of $\alpha_2$. Inverted neutrino mass ordering constraints for $\alpha_1$ and $\alpha_2$ are given in Fig.~\ref{fig:alpha1_alpha2_IO}, which shows a visible correlation between $\alpha_1$ and $\alpha_2$. A period for $\alpha_1$ is $2\pi$, whereas the period for $\alpha_2$ is $\pi$. A 1$\sigma$ regions are localized between the values of $110^\circ < \alpha_1 < 170^\circ$ and $40(220)^\circ < \alpha_2 < 60(240)^\circ$ Majorana phases.

\begin{figure*}[h]
    \centering
    \begin{subfigure}[t]{0.45\textwidth}
        \includegraphics[width=0.95\textwidth]{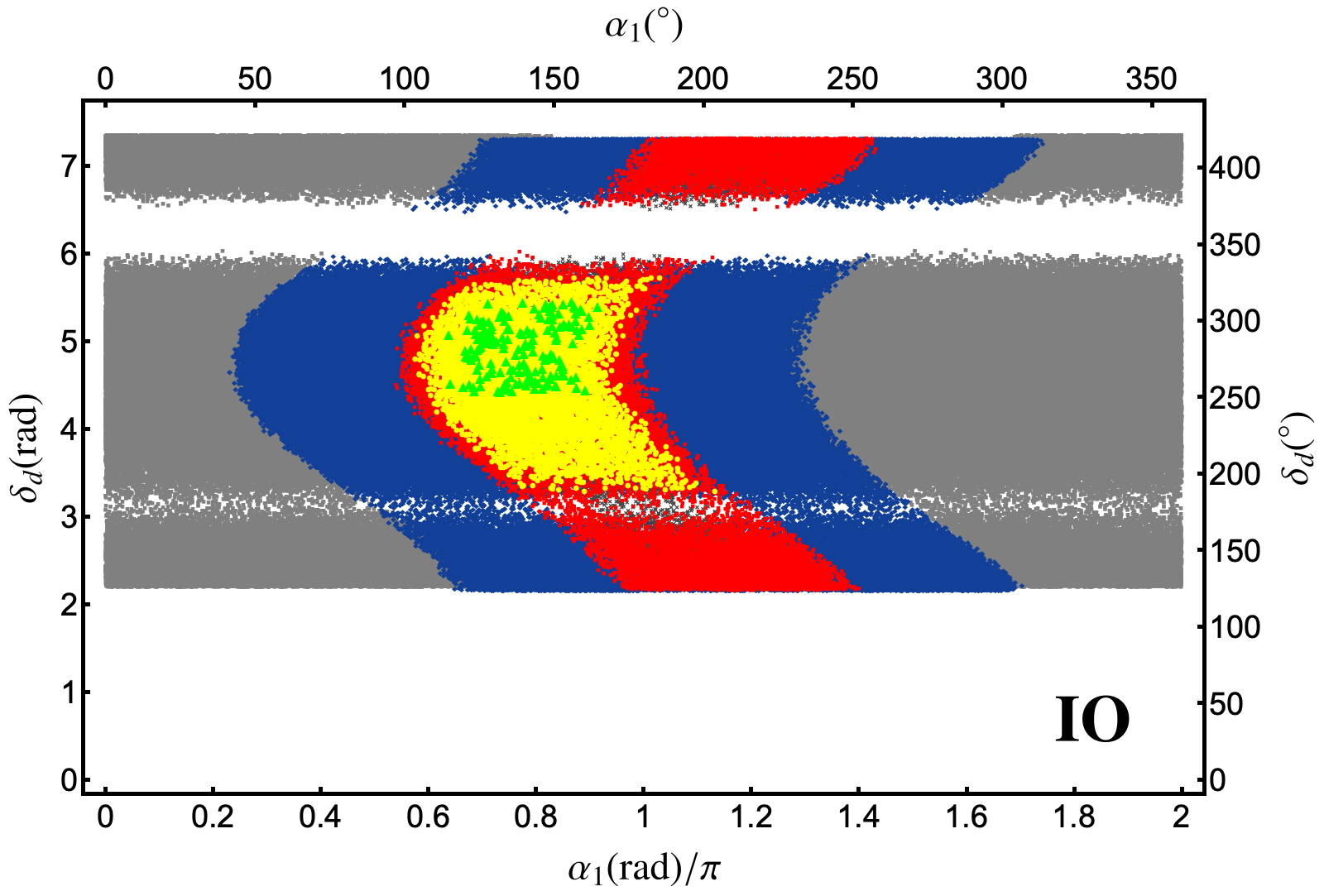}
        \caption{$\alpha_{1}$ vs $\delta_{\text{Dirac}}$ plot for inverted neutrino mass ordering.}
        \label{fig:alpha1_deltaDirac_IO}
    \end{subfigure}
    \begin{subfigure}[t]{0.45\textwidth}
        \includegraphics[width=0.95\textwidth]{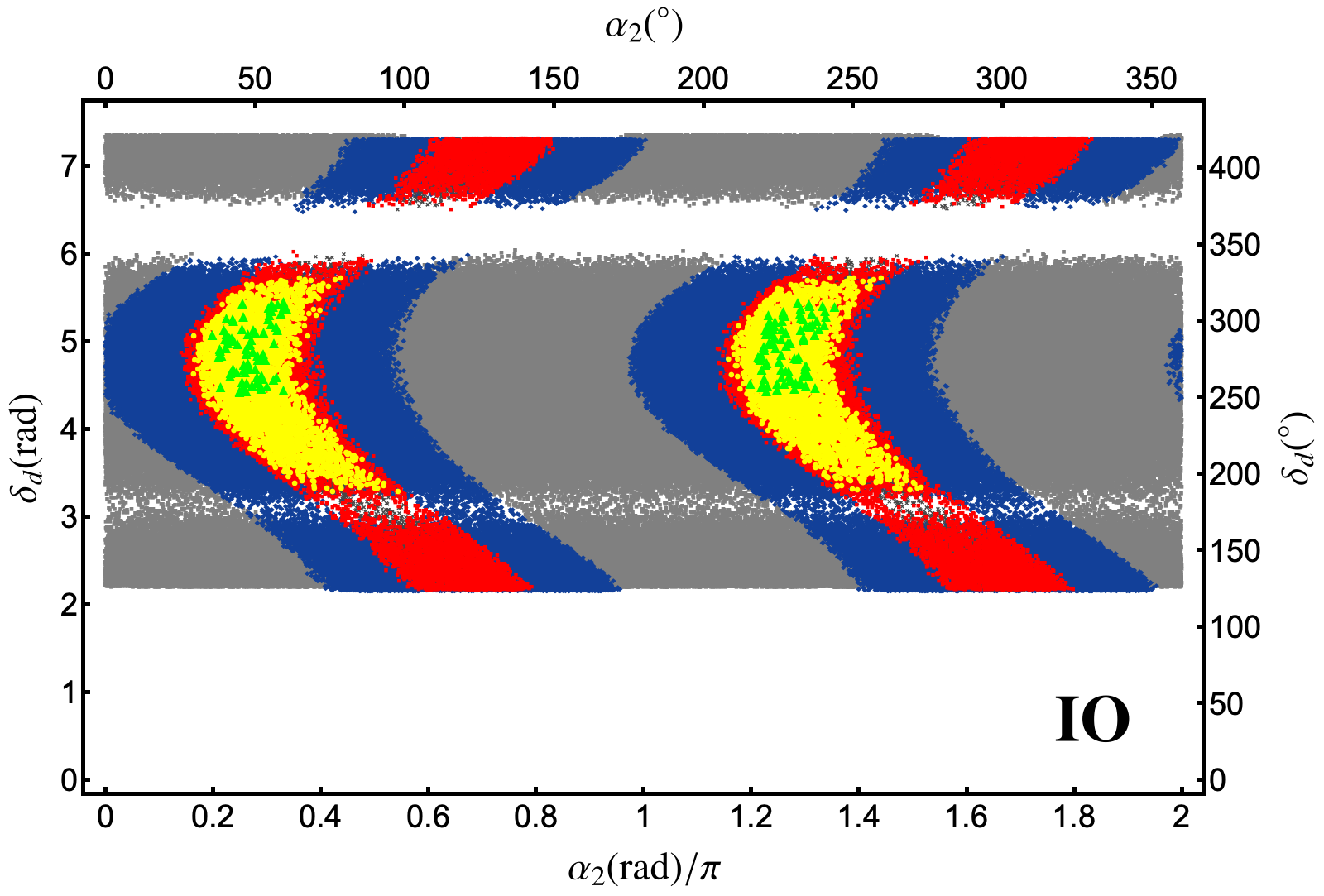}
        \caption{$\alpha_{2}$ vs $\delta_{\text{Dirac}}$ plot for inverted neutrino mass ordering.}
        \label{fig:alpha2_deltaDirac_IO}
    \end{subfigure}
    \caption{Inverted neutrino mass ordering. $\alpha_{1,2}$ vs $\delta_{\text{Dirac}}$ correlation. Color labeling is identical to that of plot in the Fig.~\ref{fig:alpha1_alpha2_IO}.}
    \label{fig:alpha12_deltaDirac_IO_plots}
\end{figure*}

%\alpha_{1,2} vs \delta_{\text{Dirac}} plots description, Fig. 2
Figs.~\ref{fig:alpha12_deltaDirac_IO_plots} shows a strong correlation between $\alpha_{1,2}$ and $\delta_{\text{Dirac}}$ phases. Even though, there are some points missing around the $\delta_{\text{Dirac}} = \pi, 2\pi$ region, for both $\alpha_{1}$ vs $\delta_{\text{Dirac}}$ and $\alpha_{2}$ vs $\delta_{\text{Dirac}}$  plots, the rest of the graphics exhibit a smooth and continuous correlation. The $1\sigma$ regions (plotted with green points) are $110^\circ < \alpha_1 < 170^\circ$, $40(220)^\circ < \alpha_2 < 60(240)^\circ$, and $240^\circ < \delta_{\text{Dirac}} < 310^\circ$ for the inverted neutrino mass ordering. As was mentioned earlier, the solution for $\alpha_2$ exhibits half of the period of the $\alpha_1$ solution.
Normal neutrino mass ordering showed no sign of correlation for $\alpha_{1,2}$ vs $\delta_{\text{Dirac}}$, therefore it has been decided to omit them here. Finally, the points $\delta_{\text{Dirac}} < 125^\circ$ are not shown in the plots of Figs.~\ref{fig:alpha1_deltaDirac_IO} and~\ref{fig:alpha2_deltaDirac_IO} because only the regions within $5\sigma$ for $\delta_{\text{Dirac}}$ were studied.

\begin{figure*}[h]
    \centering
    \begin{subfigure}[t]{0.45\textwidth}
        \includegraphics[width=0.95\textwidth]{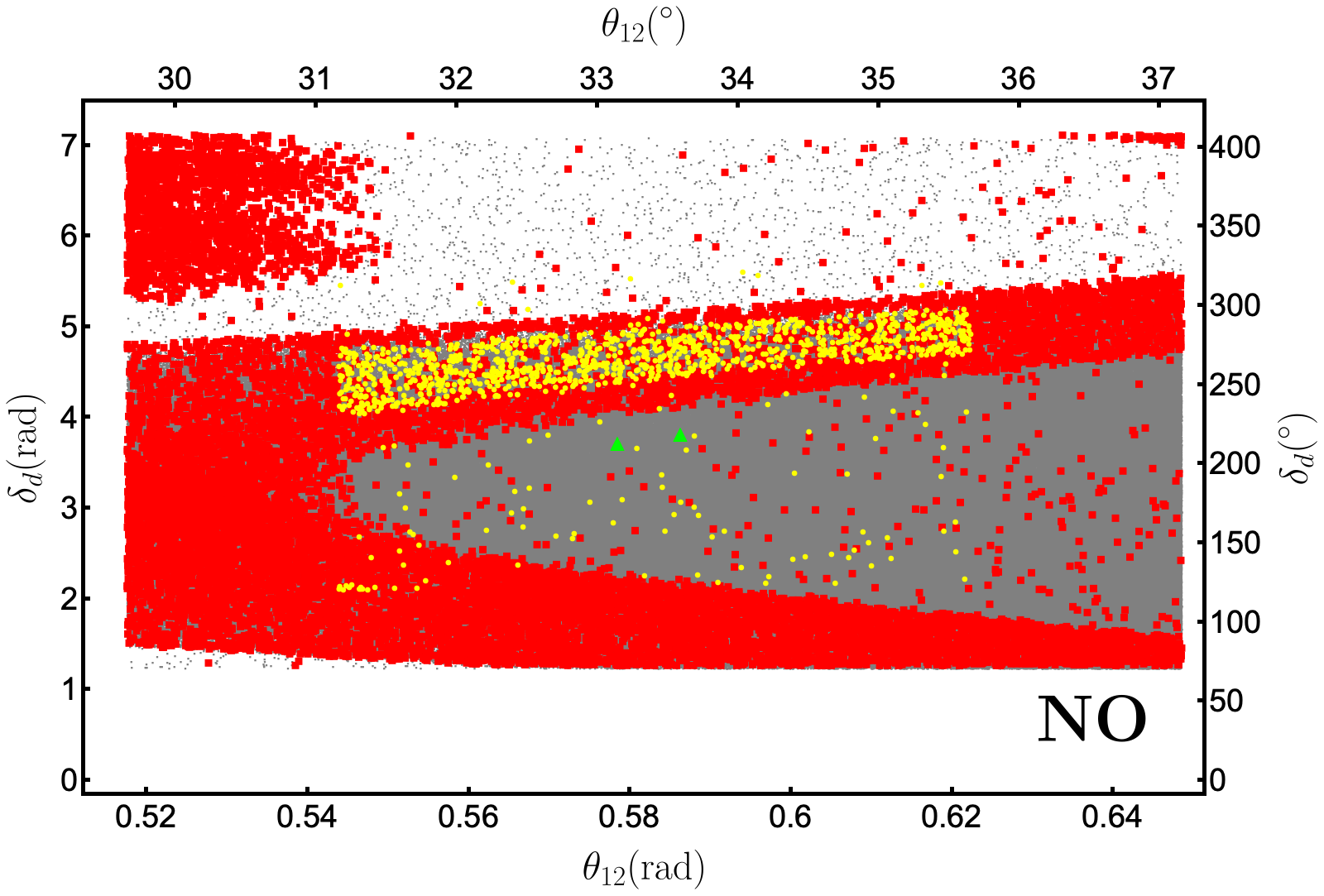}
        \caption{$\theta_{12}$ vs $\delta_{\text{Dirac}}$ plot for normal neutrino mass ordering.}
        \label{fig:theta12_deltaDirac_NO}
    \end{subfigure}
    \begin{subfigure}[t]{0.45\textwidth}
        \includegraphics[width=0.95\textwidth]{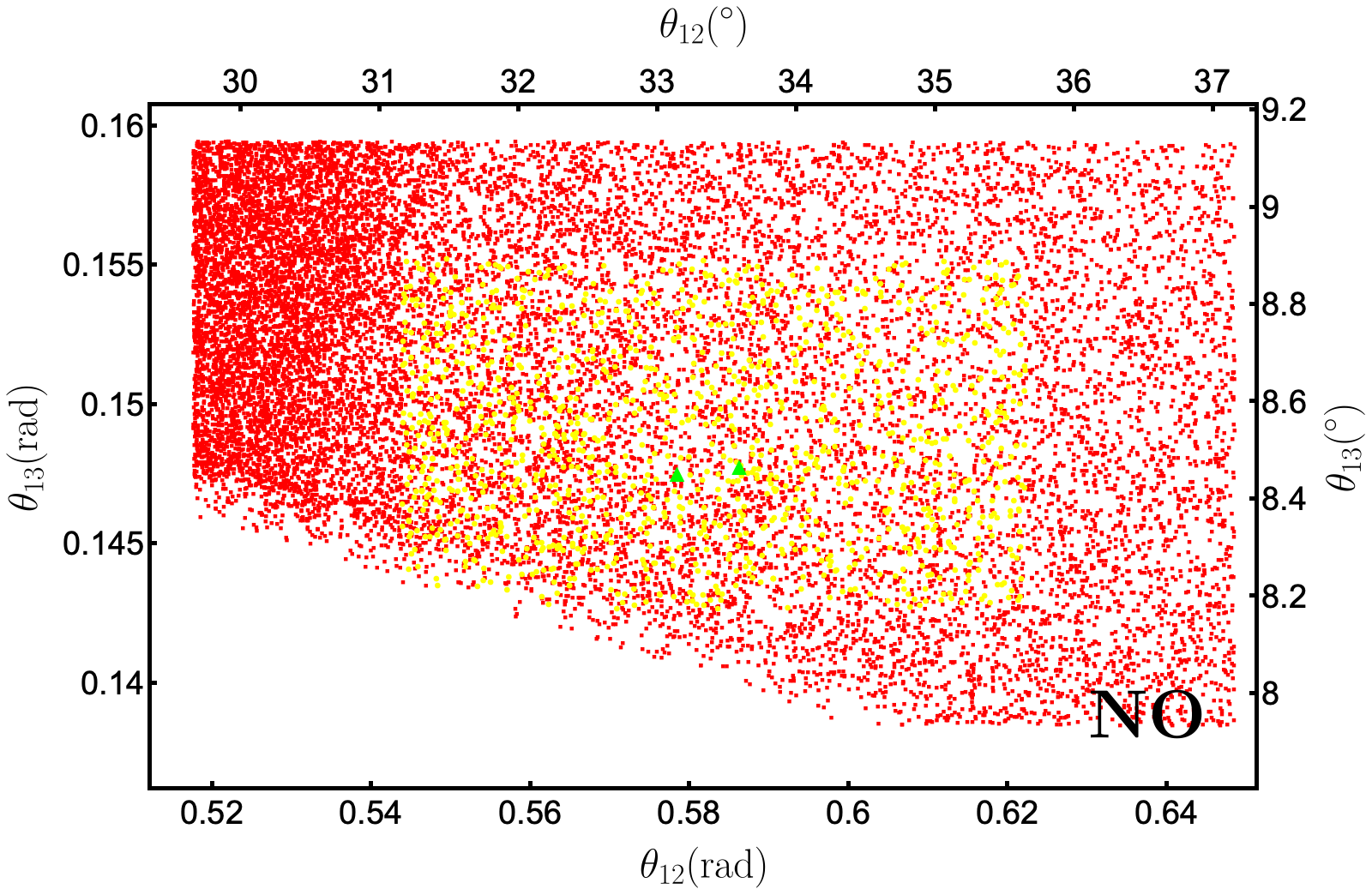}
        \caption{$\theta_{12}$ vs $\theta_{13}$ plot for normal neutrino mass ordering.}
        \label{fig:theta12_theta13_NO}
    \end{subfigure}
    \caption{Normal neutrino mass ordering. $\theta_{12}$ vs $\delta_{\text{Dirac}}$ and $\theta_{12}$ vs $\theta_{13}$ correlations. Color labeling is identical to that of plot in the Fig.~\ref{fig:alpha1_alpha2_NO}.}
    \label{fig:theta12_vs_deltaDirac_and_theta13_NO_plots}
\end{figure*}

%\theta_{12} vs \delta_{\text{Dirac} and \theta_{12} vs \theta_{13} plots description, Fig. 3
Fig.~\ref{fig:theta12_deltaDirac_NO} give a hint of weak correlation between $\theta_{12}$ and $\delta_{\text{Dirac}}$ for normal neutrino mass ordering. Even though, Fig.~\ref{fig:theta12_theta13_NO} lacks any sign of correlation between $\theta_{12}$ and $\theta_{13}$ mixing angles, nevertheless, it is important to mention that lower range of $\theta_{12}$ and simultaneously lower region of $\theta_{13}$ $5\sigma$ ranges of their experimentally measured values are missing. The corresponding inverted neutrino mass ordering plots were not included due to lack of any correlation.

\begin{figure*}[h]
    \centering
    \begin{subfigure}[t]{0.45\textwidth}
        \includegraphics[width=0.95\textwidth]{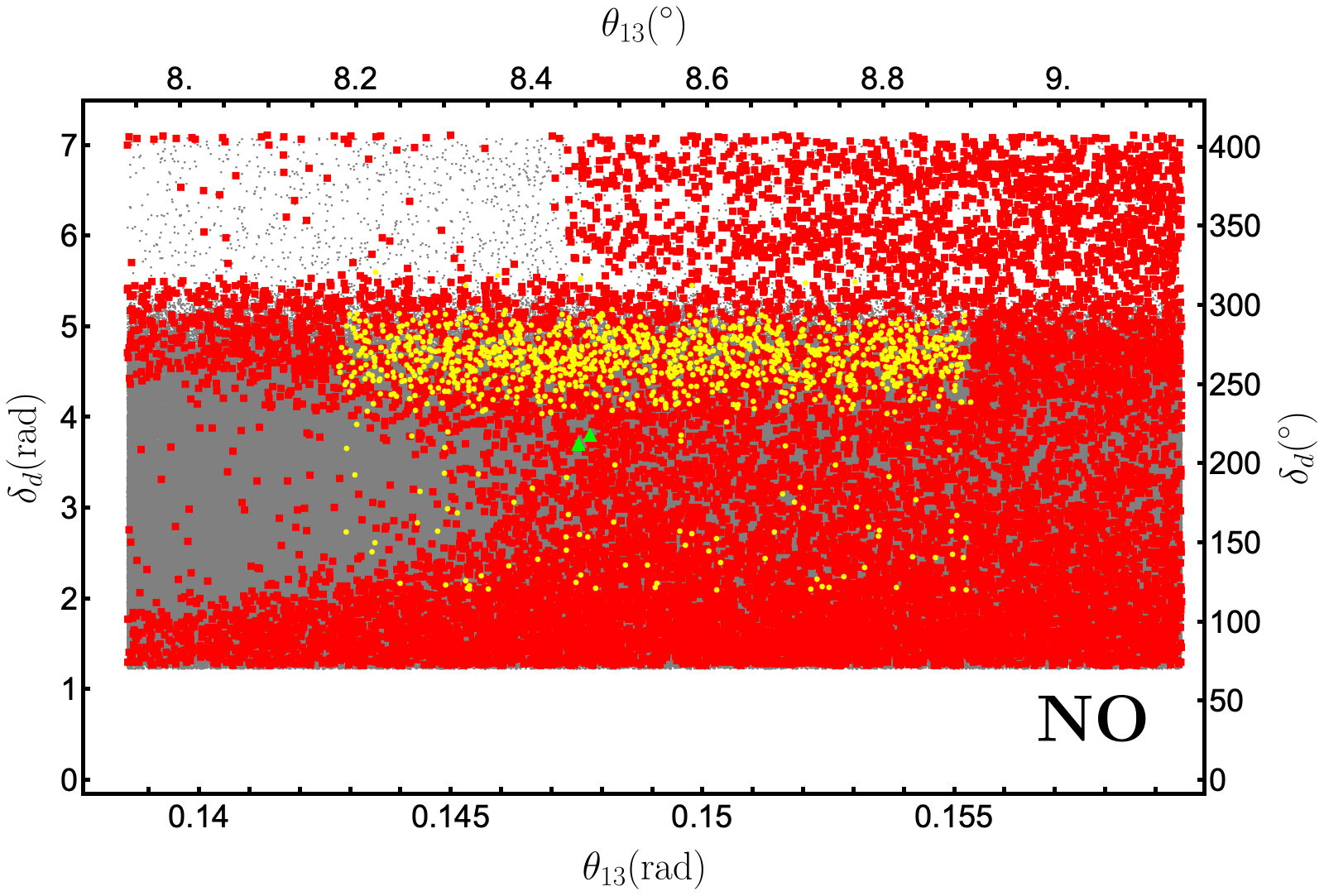}
        \caption{$\theta_{13}$ vs $\delta_{\text{Dirac}}$ plot for normal neutrino mass ordering.}
        \label{fig:theta13_deltaDirac_NO}
    \end{subfigure}
    \begin{subfigure}[t]{0.45\textwidth}
        \includegraphics[width=0.95\textwidth]{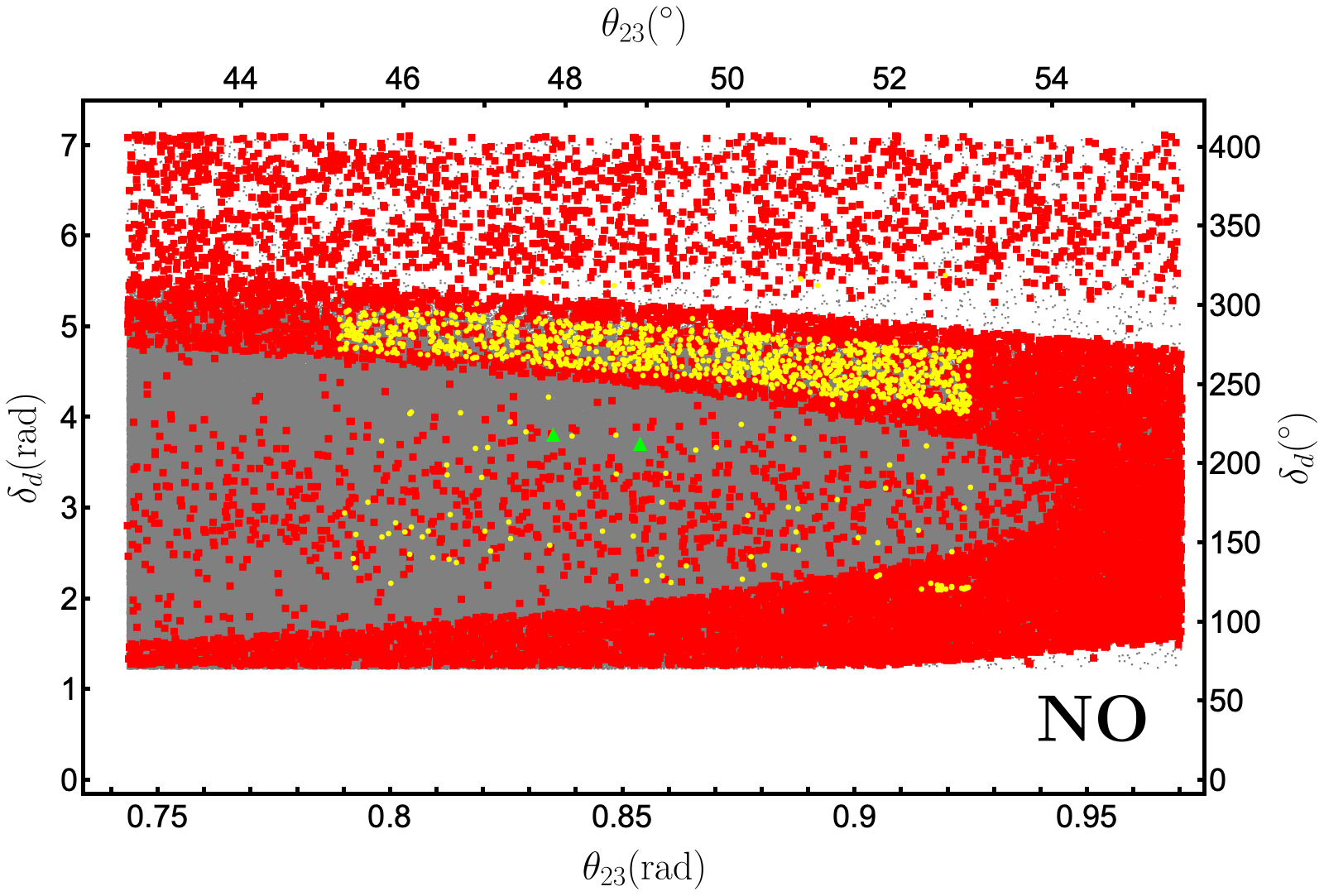}
        \caption{$\theta_{23}$ vs $\delta_{\text{Dirac}}$ plot for normal neutrino mass ordering.}
        \label{fig:theta23_deltaDirac_NO}
    \end{subfigure}
    \caption{Normal neutrino mass ordering. $\theta_{13}$ vs $\delta_{\text{Dirac}}$ and $\theta_{23}$ vs $\delta_{\text{Dirac}}$ correlations. Color labeling is identical to that of plot in the Fig.~\ref{fig:alpha1_alpha2_NO}.}
    \label{fig:theta13_and_theta23_vs_deltaDirac_NO_plots}
\end{figure*}

%\theta_{13} vs \delta_{\text{Dirac}} and \theta_{23} vs \delta_{\text{Dirac}} plots description, Fig. 4
The Fig.~\ref{fig:theta13_and_theta23_vs_deltaDirac_NO_plots} shows the results for other two lepton mixing angles for normal neutrino mass ordering. Fig.~\ref{fig:theta13_deltaDirac_NO} depicts the $\theta_{13}$ vs $\delta_{\text{Dirac}}$ correlation and Fig.~\ref{fig:theta23_deltaDirac_NO} depicts the $\theta_{23}$ vs $\delta_{\text{Dirac}}$ correlation for normal neutrino mass ordering.  The corresponding inverted neutrino mass ordering plots were not included due to lack of any correlation.

\begin{figure*}[h]
    \centering
    \begin{subfigure}[t]{0.45\textwidth}
        \includegraphics[width=0.95\textwidth]{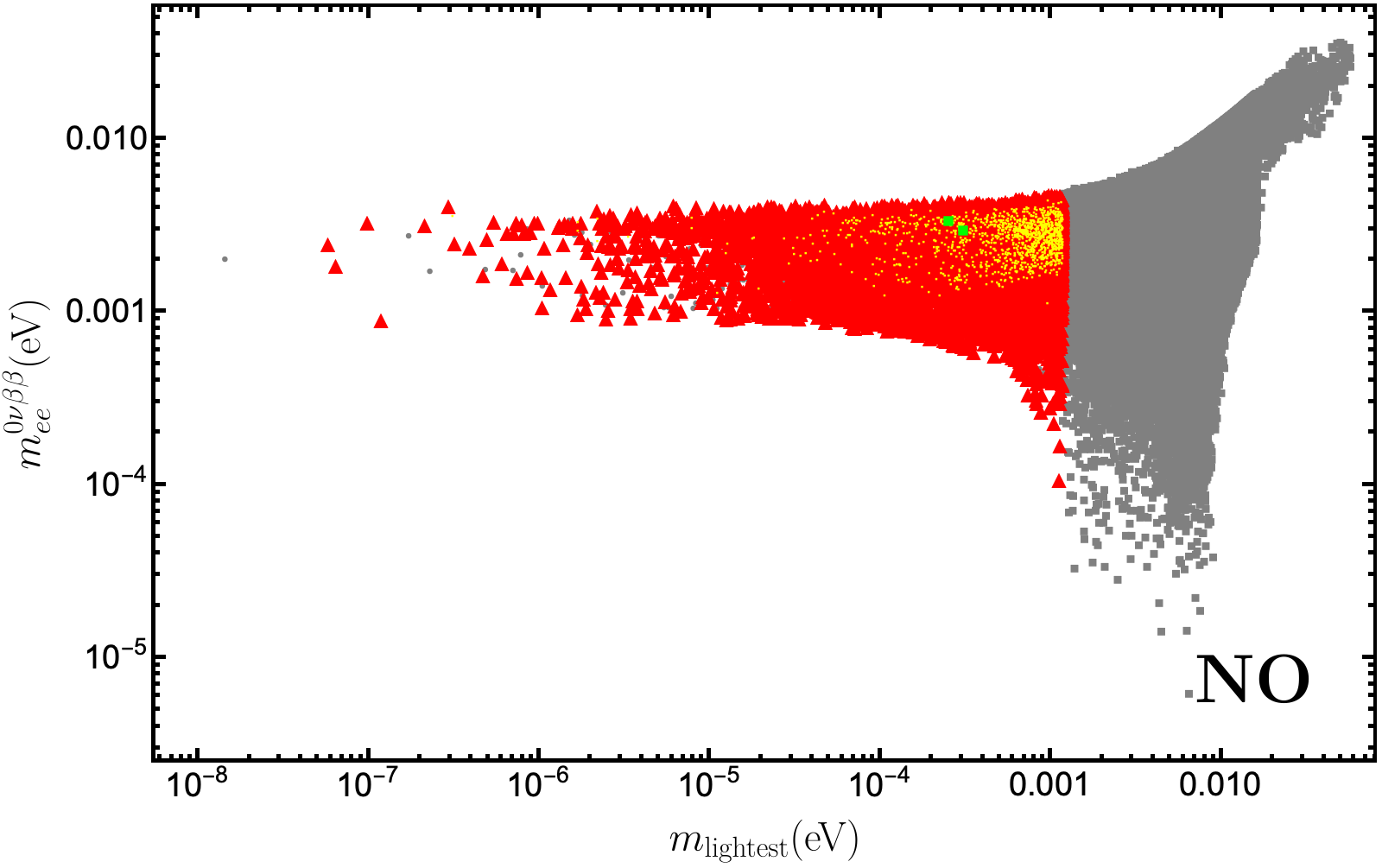}
        \caption{$m_{\text{lightest}}$ vs $m_{ee}^{0\nu\beta\beta}$ plot for normal neutrino mass ordering.}
        \label{fig:mlightest_0nubetabeta_NO}
    \end{subfigure}
    \begin{subfigure}[t]{0.45\textwidth}
        \includegraphics[width=0.95\textwidth]{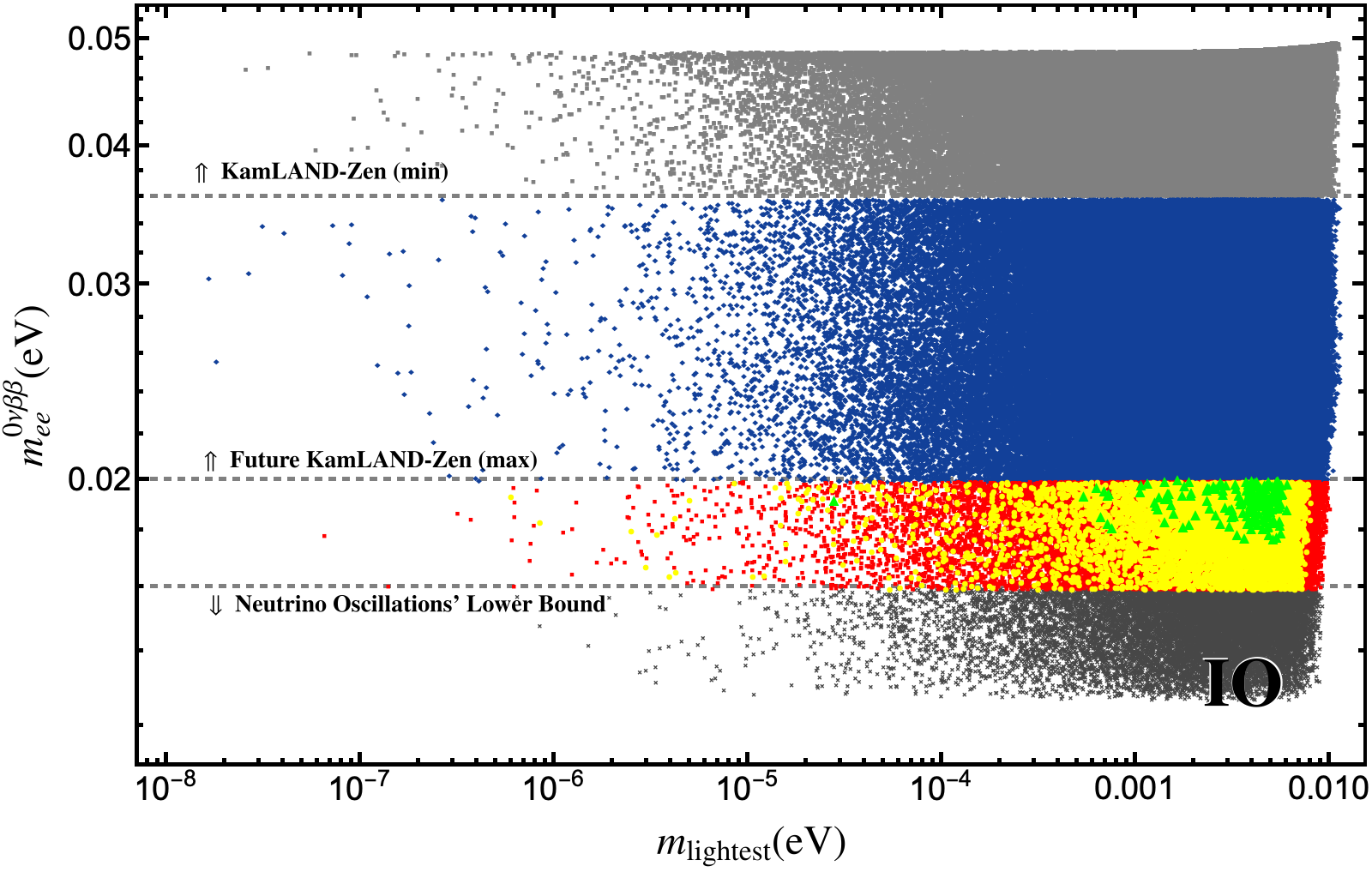}
        \caption{$m_{\text{lightest}}$ vs $m_{ee}^{0\nu\beta\beta}$ plot for inverted neutrino mass ordering.}
        \label{fig:mlightest_0nubetabeta_IO}
    \end{subfigure}
    \caption{$m_{\text{lightest}}$ vs $m_{ee}^{0\nu\beta\beta}$ correlations for normal and inverted neutrino mass ordering. Color labeling is identical to that of plots in the Fig.~\ref{fig:alpha1_alpha2_plots}.}
    \label{fig:mlightest_vs_0nubetabeta_plots}
\end{figure*}

%m_{\text{lightest}} vs m_{ee}^{0\nu\beta\beta} plots description, Fig. 5
Plots in Fig.~\ref{fig:mlightest_vs_0nubetabeta_plots} show the survival region for neutrinoless double decay prediction for both normal and inverted ordering. As is seen from plot in Fig.~\ref{fig:mlightest_0nubetabeta_NO} that only a small region can satisfy the $m_{\nu}^{13} + m_{\nu}^{22} + m_{\nu}^{31} = 0$ neutrino mass selection rule and lepton mixing angles within $1\sigma$ bound, $m_{\text{lightest}} \approx 3 - 4 \times 10^{-4}$~eV and $m_{ee}^{0\nu\beta\beta} \approx 3 \times 10^{-3}$~eV for normal neutrino mass ordering. The survived regions are well below the current ($m_{ee}^{0\nu\beta\beta} < 0.036 - 0.156$~eV) and future KamLAND-Zen ($m_{ee}^{0\nu\beta\beta} < 0.010 - 0.020$~eV) sensitivity regions. On the other hand, the situation for inverted neutrino mass ordering reversed (See Fig.~\ref{fig:mlightest_0nubetabeta_IO}). The $1\sigma$ survival region $m_{\text{lightest}} \approx 2 \times 10^{-5} - 7 \times 10^{-3}$~eV is squeezed between the neutrino oscillations' lower bound and future KamLAND-Zen sensitivity bound. 

\begin{figure*}[h]
    \centering
    \begin{subfigure}[t]{0.45\textwidth}
        \includegraphics[width=0.95\textwidth]{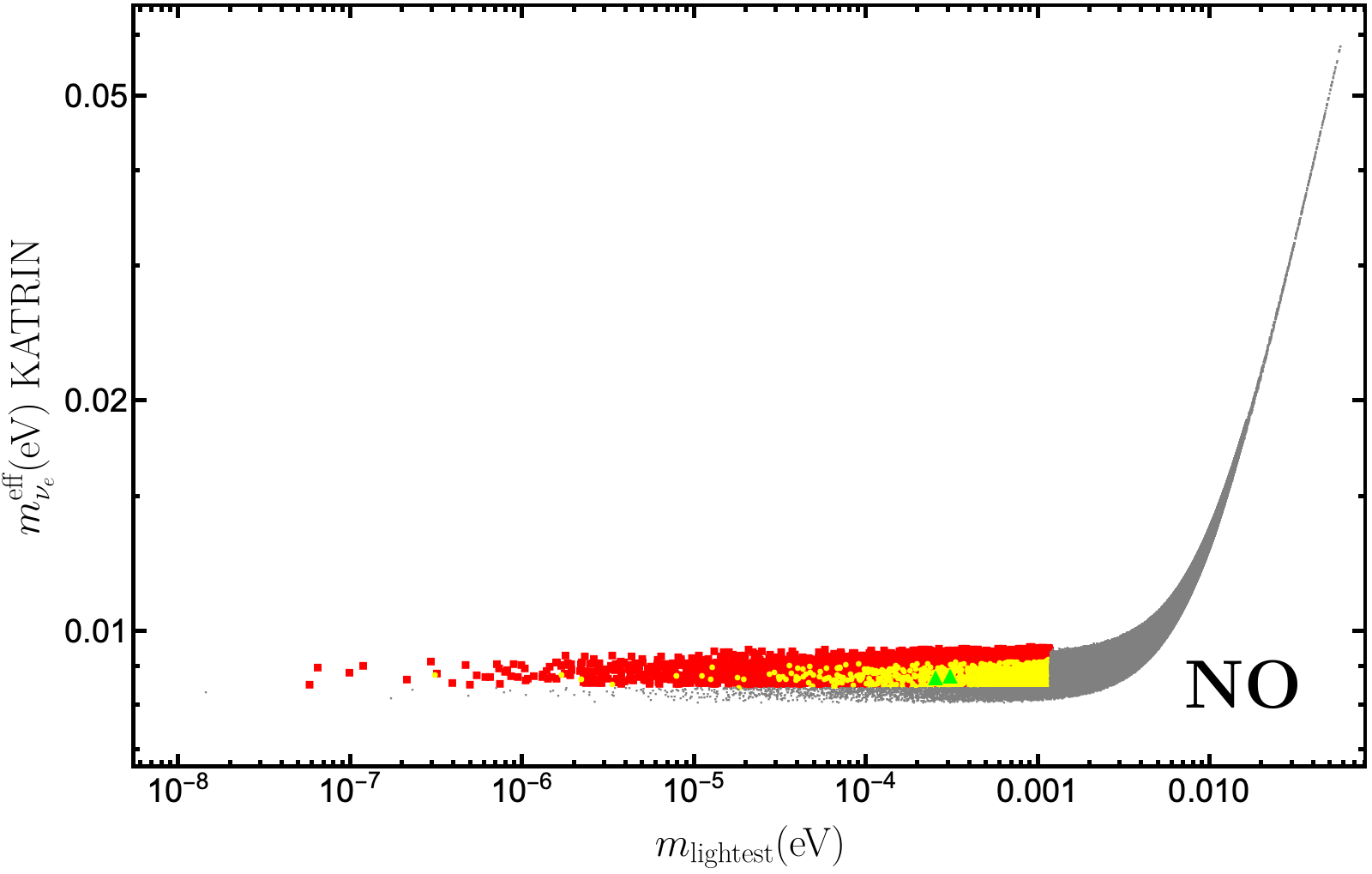}
        \caption{$m_{\text{lightest}}$ vs $m_{\nu_e}^{\text{eff}}$ plot for normal neutrino mass ordering.}
        \label{fig:mlightest_mnueeff_NO}
    \end{subfigure}
    \begin{subfigure}[t]{0.45\textwidth}
        \includegraphics[width=0.95\textwidth]{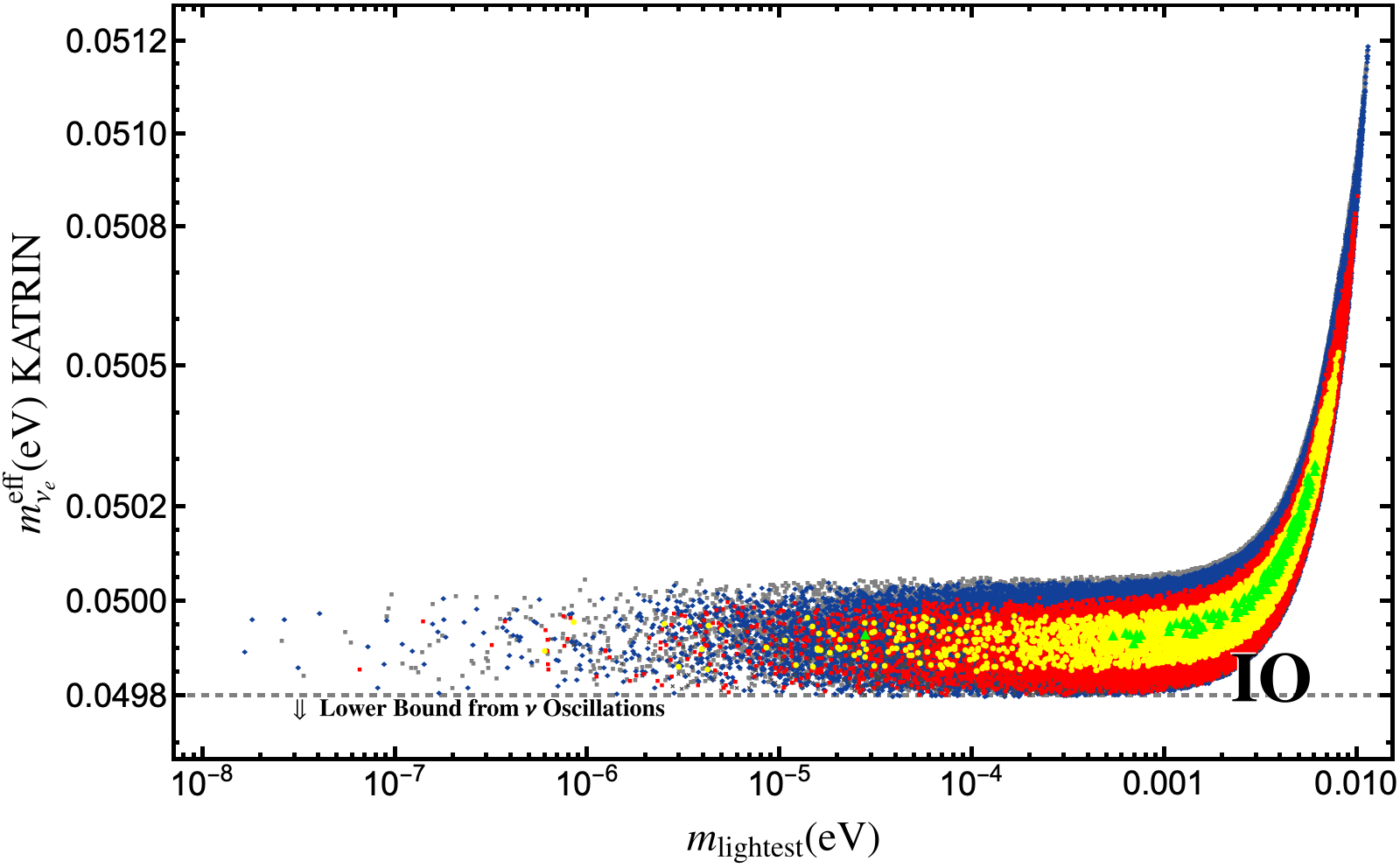}
        \caption{$m_{\text{lightest}}$ vs $m_{\nu_e}^{\text{eff}}$ plot for inverted neutrino mass ordering.}
        \label{fig:mlightest_mnueeff_IO}
    \end{subfigure}
    \caption{$m_{\text{lightest}}$ vs $m_{\nu_e}^{\text{eff}}$ correlations for normal and inverted neutrino mass ordering. Color labeling is identical to that of plots in the Fig.~\ref{fig:alpha1_alpha2_plots}.}
    \label{fig:mlightest_vs_mnueeff_plots}
\end{figure*}

%m_{\text{lightest}} vs m_{\nu_e}^{\text{eff}} plots description, Fig. 6
The survival regions for $m_{\nu_e}^{\text{eff}}$, subject to KATRIN experiment measurements, are well below the current and expected sensitivity regions ($m_{\nu_e}^{\text{eff}} < 0.2 - 0.45$~eV~\cite{KATRIN:2024cdt,KATRIN:2021uub,Drexlin:2013lha}). The is shown in Fig.~\ref{fig:mlightest_vs_mnueeff_plots} for normal (See Fig.~\ref{fig:mlightest_mnueeff_NO}) and inverted (See Fig.~\ref{fig:mlightest_mnueeff_IO}) neutrino mass orderings. Obtained constraints on $m_{\text{lightest}} \approx 3 - 4 \times 10^{-4}$~eV for normal neutrino mass ordering and $m_{\text{lightest}} \approx 2 \times 10^{-5} - 7 \times 10^{-3}$~eV for inverted neutrino mass ordering are the same as in the previous paragraph.

\begin{figure*}[h]
    \centering
    \begin{subfigure}[t]{0.45\textwidth}
        \includegraphics[width=0.95\textwidth]{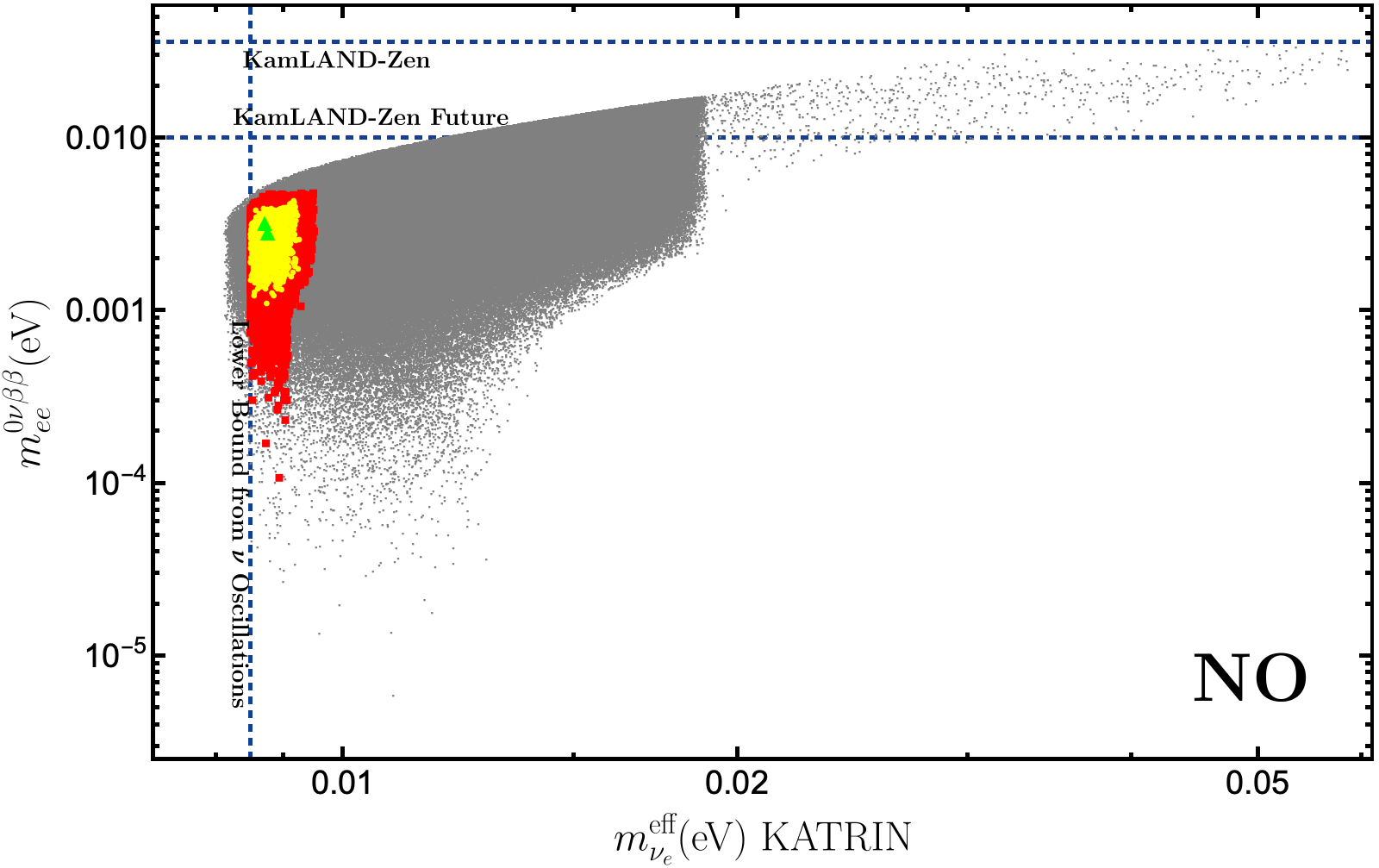}
        \caption{$m_{\nu_e}^{\text{eff}}$ vs $m_{ee}^{0\nu\beta\beta}$ plot for normal neutrino mass ordering.}
        \label{fig:mnueeff_0nubetabeta_NO}
    \end{subfigure}
    \begin{subfigure}[t]{0.45\textwidth}
        \includegraphics[width=0.95\textwidth]{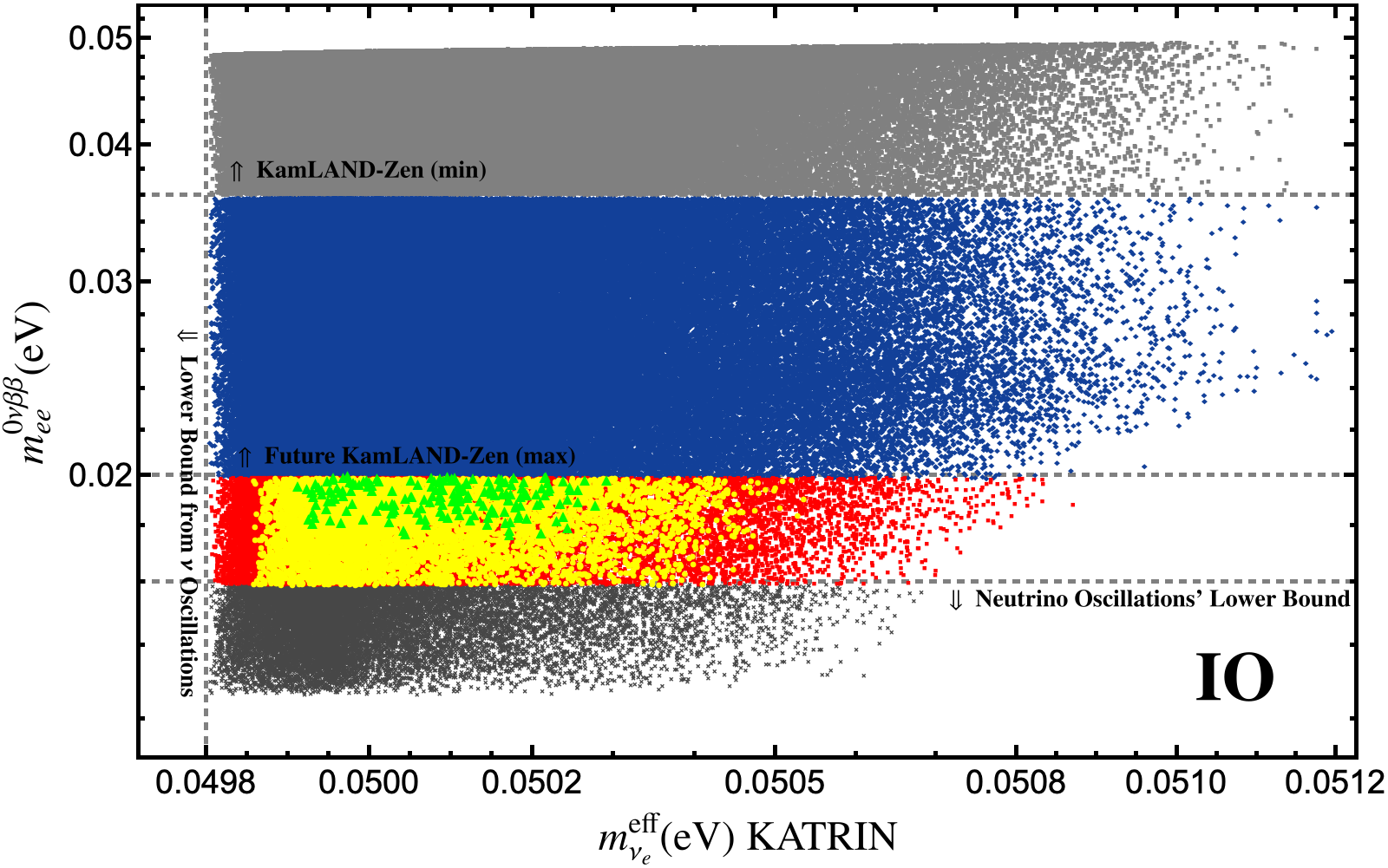}
        \caption{$m_{\nu_e}^{\text{eff}}$ vs $m_{ee}^{0\nu\beta\beta}$ plot for inverted neutrino mass ordering.}
        \label{fig:mnueeff_0nubetabeta_IO}
    \end{subfigure}
    \caption{$m_{\nu_e}^{\text{eff}}$ vs $m_{ee}^{0\nu\beta\beta}$ correlations for normal and inverted neutrino mass ordering. Color labeling is identical to that of plots in the Fig.~\ref{fig:alpha1_alpha2_plots}.}
    \label{fig:mnueeff_0nubetabeta_plots}
\end{figure*}

%m_{\nu_e}^{\text{eff}} vs m_{ee}^{0\nu\beta\beta} plots description, Fig. 7
Plots in Fig.~\ref{fig:mnueeff_0nubetabeta_plots} show the survival regions and experimental constraints for $m_{\nu_e}^{\text{eff}}$ vs $m_{ee}^{0\nu\beta\beta}$ observable space. The figure also shows the experimental bounds, current and future, for both normal (See Fig.~\ref{fig:mnueeff_0nubetabeta_NO}) and inverted (See Fig.~\ref{fig:mnueeff_0nubetabeta_IO}) neutrino mass ordering. The bounds are the same as mentioned in the preceding paragraphs.

\begin{figure*}[h]
    \centering
    \begin{subfigure}[t]{0.45\textwidth}
        \includegraphics[width=0.95\textwidth]{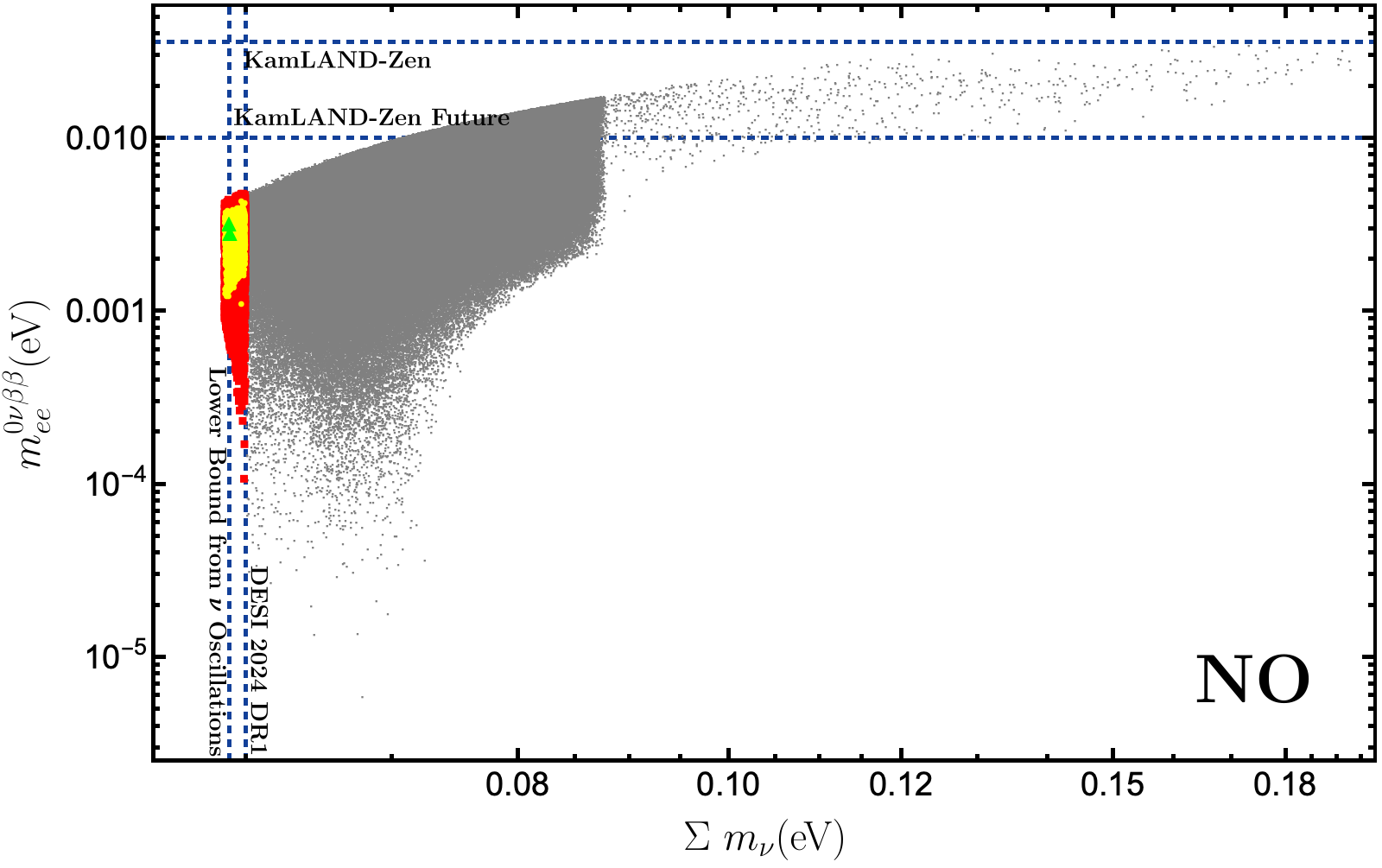}
        \caption{$\sum m_{\nu}$ vs $m_{ee}^{0\nu\beta\beta}$ plot for normal neutrino mass ordering.}
        \label{fig:summnu_0nubetabeta_NO}
    \end{subfigure}
    \begin{subfigure}[t]{0.45\textwidth}
        \includegraphics[width=0.95\textwidth]{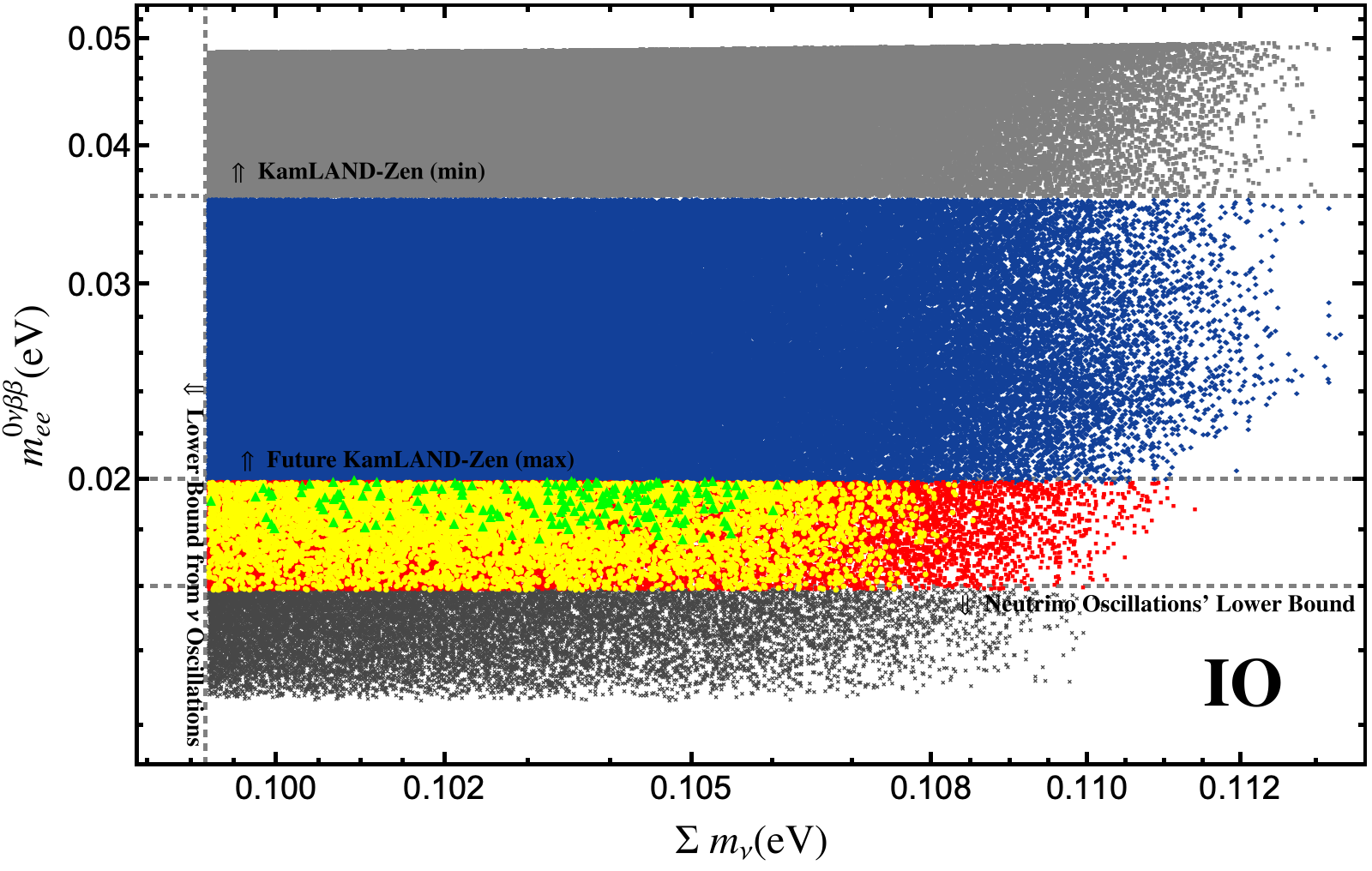}
        \caption{$\sum m_{\nu}$ vs $m_{ee}^{0\nu\beta\beta}$ plot for inverted neutrino mass ordering.}
        \label{fig:summnu_0nubetabeta_IO}
    \end{subfigure}
    \caption{$\sum m_{\nu}$ vs $m_{ee}^{0\nu\beta\beta}$ correlations for normal and inverted neutrino mass ordering. Color labeling is identical to that of plots in the Fig.~\ref{fig:alpha1_alpha2_plots}.}
    \label{fig:summnu_0nubetabeta_plots}
\end{figure*}

%\sum m_{\nu} vs m_{ee}^{0\nu\beta\beta} plots description, Fig. 8
Fig.~\ref{fig:summnu_0nubetabeta_plots} shows the survival regions and experimental constraints for $\sum m_{\nu}$ vs $m_{ee}^{0\nu\beta\beta}$ observable space. For normal neutrino mass ordering (See Fig.~\ref{fig:summnu_0nubetabeta_NO}) the most up-to-date $\sum m_{\nu} < 0.06$~eV~\cite{Wang:2024hen} cosmological bound is considered.  These cosmological bound is a combination of cosmic microwave background (CMB), DESI 2024 Baryon Acoustic Oscillations (BAO), Supernovae Ia, Gamma Ray Bursts (GRB), etc. On the contrary side, for inverted neutrino mass ordering (See Fig.~\ref{fig:summnu_0nubetabeta_NO}) the $\sum m_{\nu} < 0.072 (0.113)$~eV~\cite{DESI:2024mwx,DESI:2024hhd} for $\sum m_{\nu} > 0 (0.059)$~eV prior from a recent DESI 2024 DR1 cosmological constraint is considered. For the $\sum m_{\nu} < 0.072$~eV  value no surviving points left, since this bound is below the lower bound from the neutrino oscillations for inverted neutrino mass ordering. Whereas, for the $\sum m_{\nu} < 0.113$~eV cosmological bound all points within $5\sigma$ range of lepton mixing angles, that also satisfy the $m_{\nu}^{13} + m_{\nu}^{22} + m_{\nu}^{31} = 0$ neutrino mass selection rule do survive. Finally, all points in Fig.~\ref{fig:summnu_0nubetabeta_plots} are well below the $\sum m_{\nu} < 0.265 (0.218)$~eV~\cite{Planck:2018vyg} bounds from PLANCK 2018 CMB Results with(out) CMB lensing.

\begin{figure*}[h]
    \centering
    \begin{subfigure}[t]{0.45\textwidth}
        \includegraphics[width=0.95\textwidth]{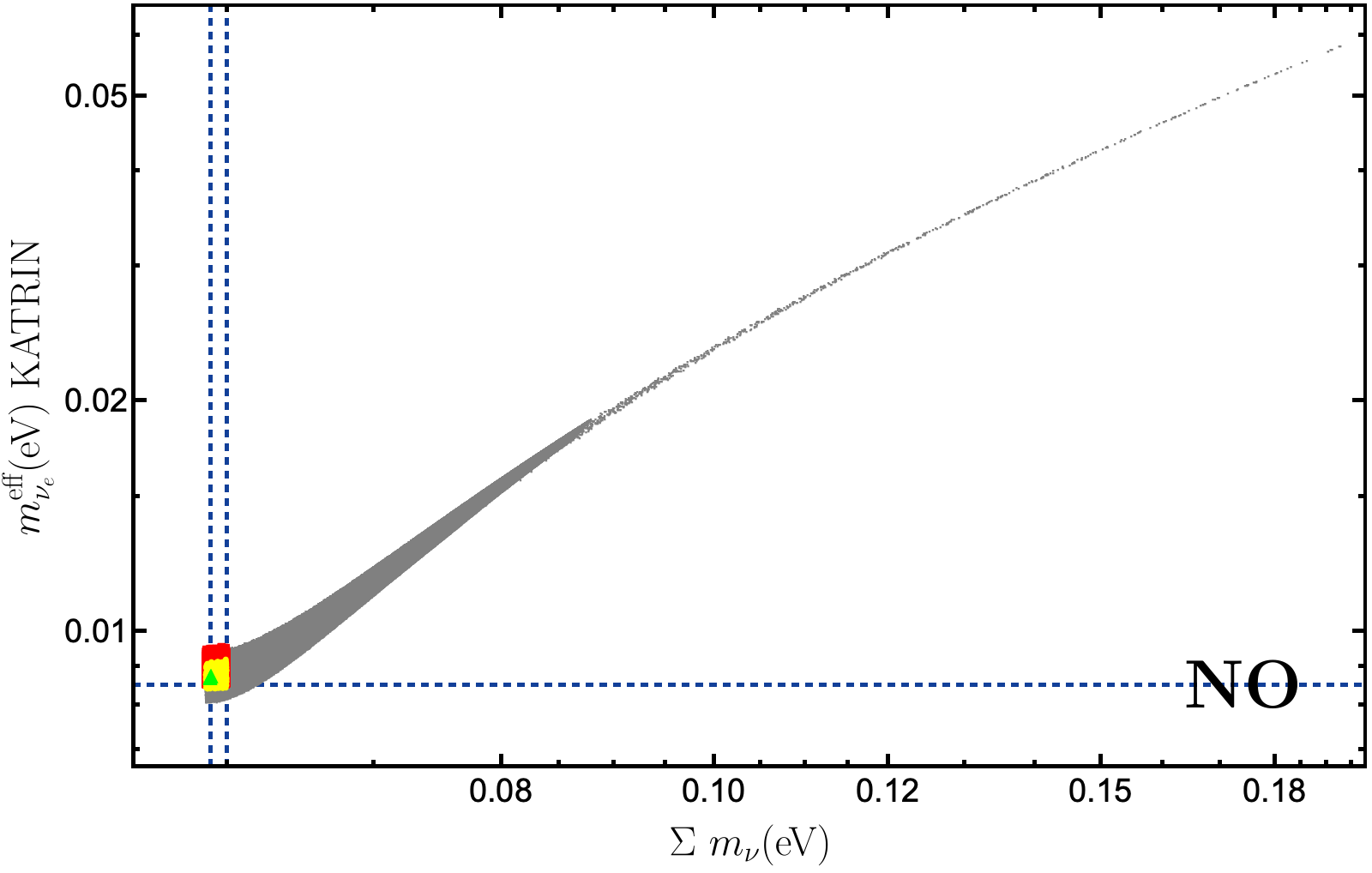}
        \caption{$\sum m_{\nu}$ vs $m_{\nu_e}^{\text{eff}}$ plot for normal neutrino mass ordering.}
        \label{fig:summnu_mnueeff_NO}
    \end{subfigure}
    \begin{subfigure}[t]{0.45\textwidth}
        \includegraphics[width=0.95\textwidth]{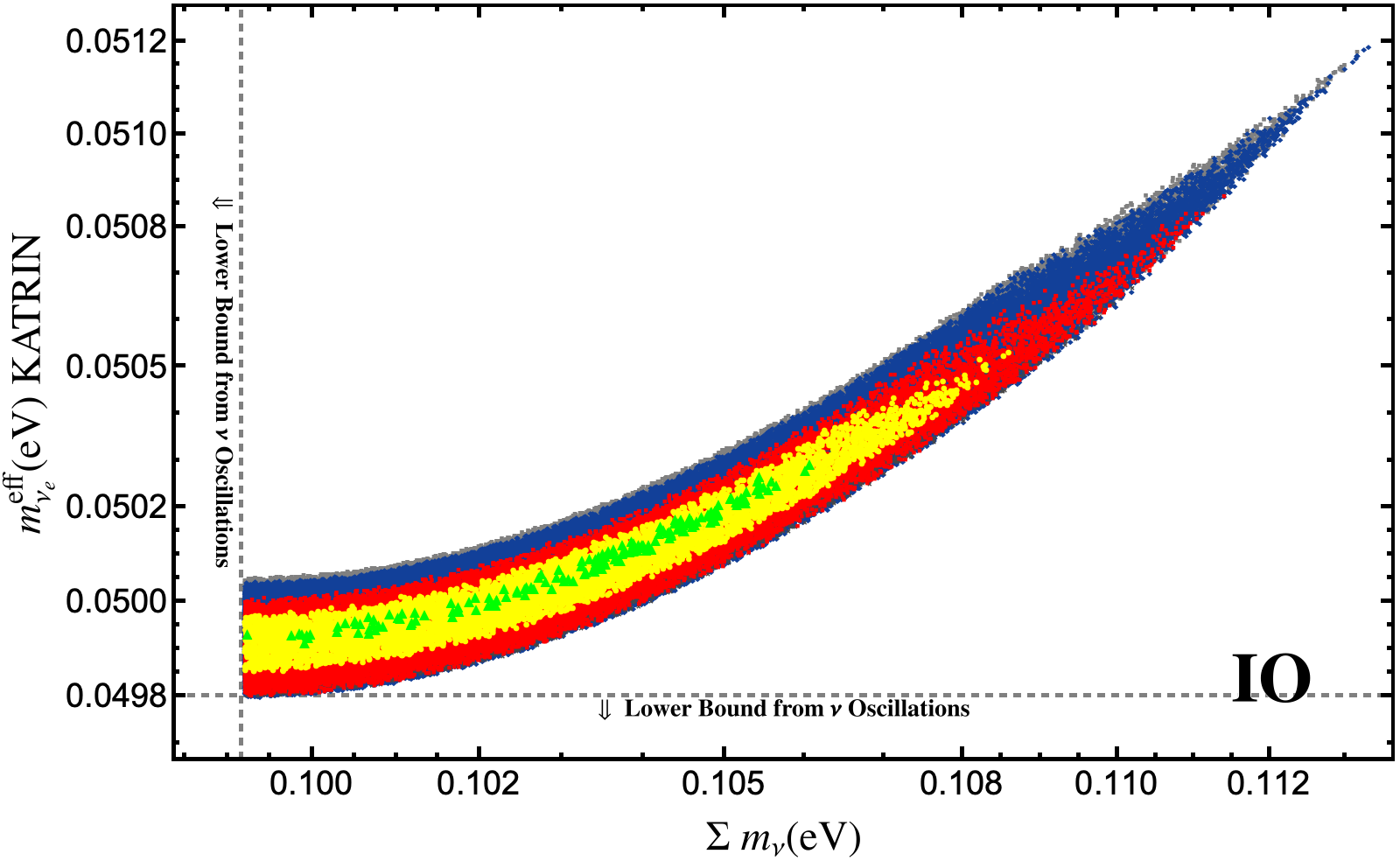}
        \caption{$\sum m_{\nu}$ vs $m_{\nu_e}^{\text{eff}}$ plot for inverted neutrino mass ordering.}
        \label{fig:summnu_mnueeff_IO}
    \end{subfigure}
    \caption{$\sum m_{\nu}$ vs $m_{\nu_e}^{\text{eff}}$ correlations for normal and inverted neutrino mass ordering. Color labeling is identical to that of plots in the Fig.~\ref{fig:alpha1_alpha2_plots}.}
    \label{fig:summnu_mnueeff_plots}
\end{figure*}

%\sum m_{\nu} vs m_{\nu_e}^{\text{eff}} plots description, Fig. 9
Next, Fig.~\ref{fig:summnu_mnueeff_plots} shows the correlation and experimental bounds for $\sum m_{\nu}$ vs $m_{\nu_e}^{\text{eff}}$ observable space. For normal neutrino mass ordering (See Fig.~\ref{fig:summnu_mnueeff_NO}), the cosmological (DESI 2024 DR1), neutrino oscillation, and KATRIN experimental bounds are considered. Whereas, for inverted neutrino mass ordering (See Fig.~\ref{fig:summnu_mnueeff_IO}), the cosmological ($\sum m_{\nu} < 0.113$~eV, DESI 2024 DR1), neutrino oscillation, and KATRIN experimental bounds are considered. Among the stringiest bounds in our study case are the neutrino oscillations' bounds and DESI 2024 DR1 constraint.

\begin{figure*}[h]
    \centering
    \begin{subfigure}[t]{0.45\textwidth}
        \includegraphics[width=0.95\textwidth]{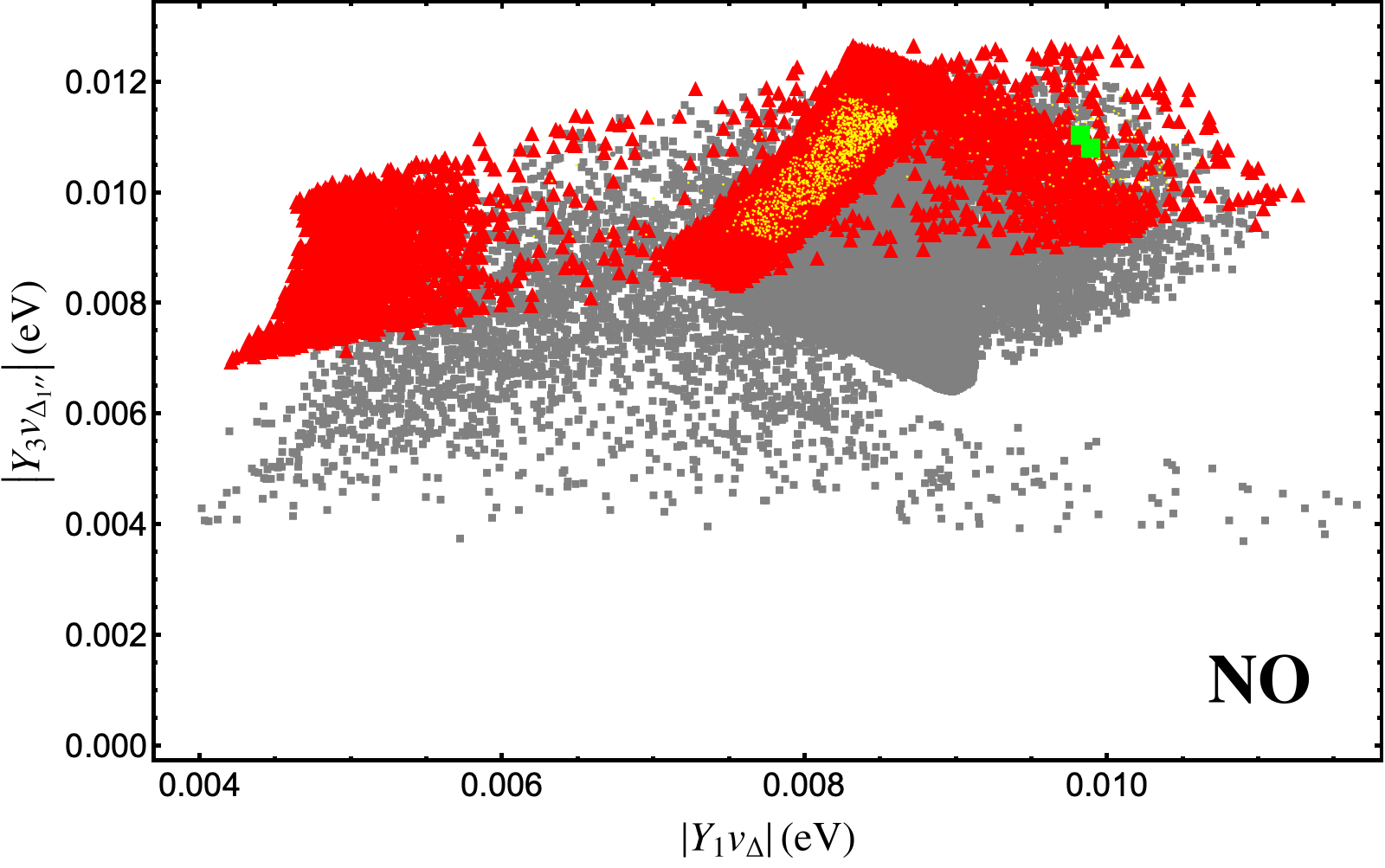}
        \caption{$\left|Y_1 v_{\Delta}\right|$ vs $\left|Y_3 v_{\Delta_{1^{\prime\prime}}}\right|$ plot for normal neutrino mass ordering.}
        \label{fig:absY1_absY3_NO}
    \end{subfigure}
    \begin{subfigure}[t]{0.45\textwidth}
        \includegraphics[width=0.95\textwidth]{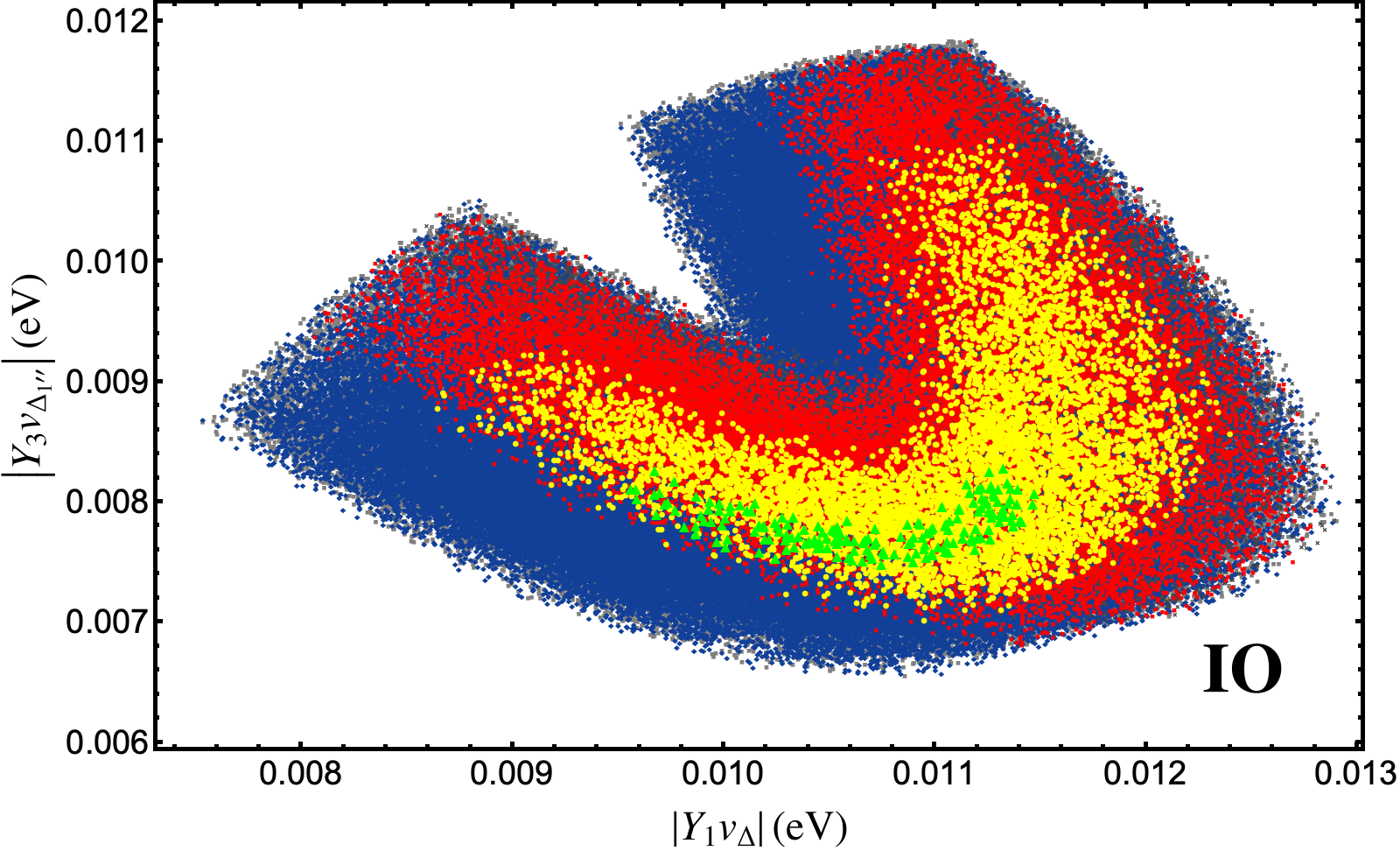}
        \caption{$\left|Y_1 v_{\Delta}\right|$ vs $\left|Y_3 v_{\Delta_{1^{\prime\prime}}}\right|$ plot for inverted neutrino mass ordering.}
        \label{fig:absY1_absY3_IO}
    \end{subfigure}
    \caption{$\left|Y_1 v_{\Delta}\right|$ vs $\left|Y_3 v_{\Delta_{1^{\prime\prime}}}\right|$ correlations for normal and inverted neutrino mass ordering. Color labeling is identical to that of plots in the Fig.~\ref{fig:alpha1_alpha2_plots}.}
    \label{fig:absY1_absY3_plots}
\end{figure*}

\begin{figure*}[h]
    \centering
    \begin{subfigure}[t]{0.45\textwidth}
        \includegraphics[width=0.95\textwidth]{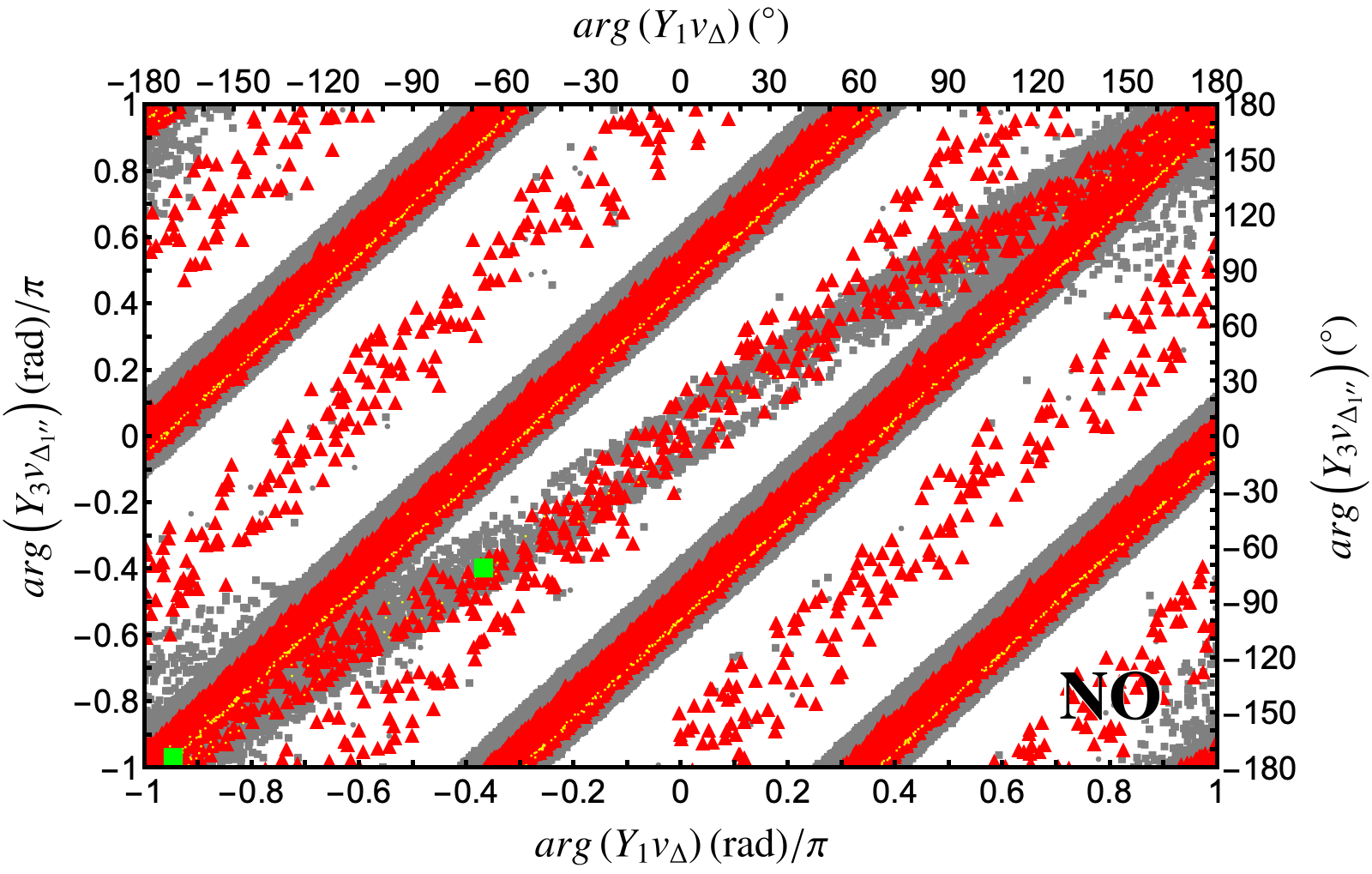}
        \caption{$arg\left(Y_1 v_{\Delta}\right)$ vs $arg\left(Y_3 v_{\Delta_{1^{\prime\prime}}}\right)$ plot for normal neutrino mass ordering.}
        \label{fig:argY1_argY3_NO}
    \end{subfigure}
    \begin{subfigure}[t]{0.45\textwidth}
        \includegraphics[width=0.95\textwidth]{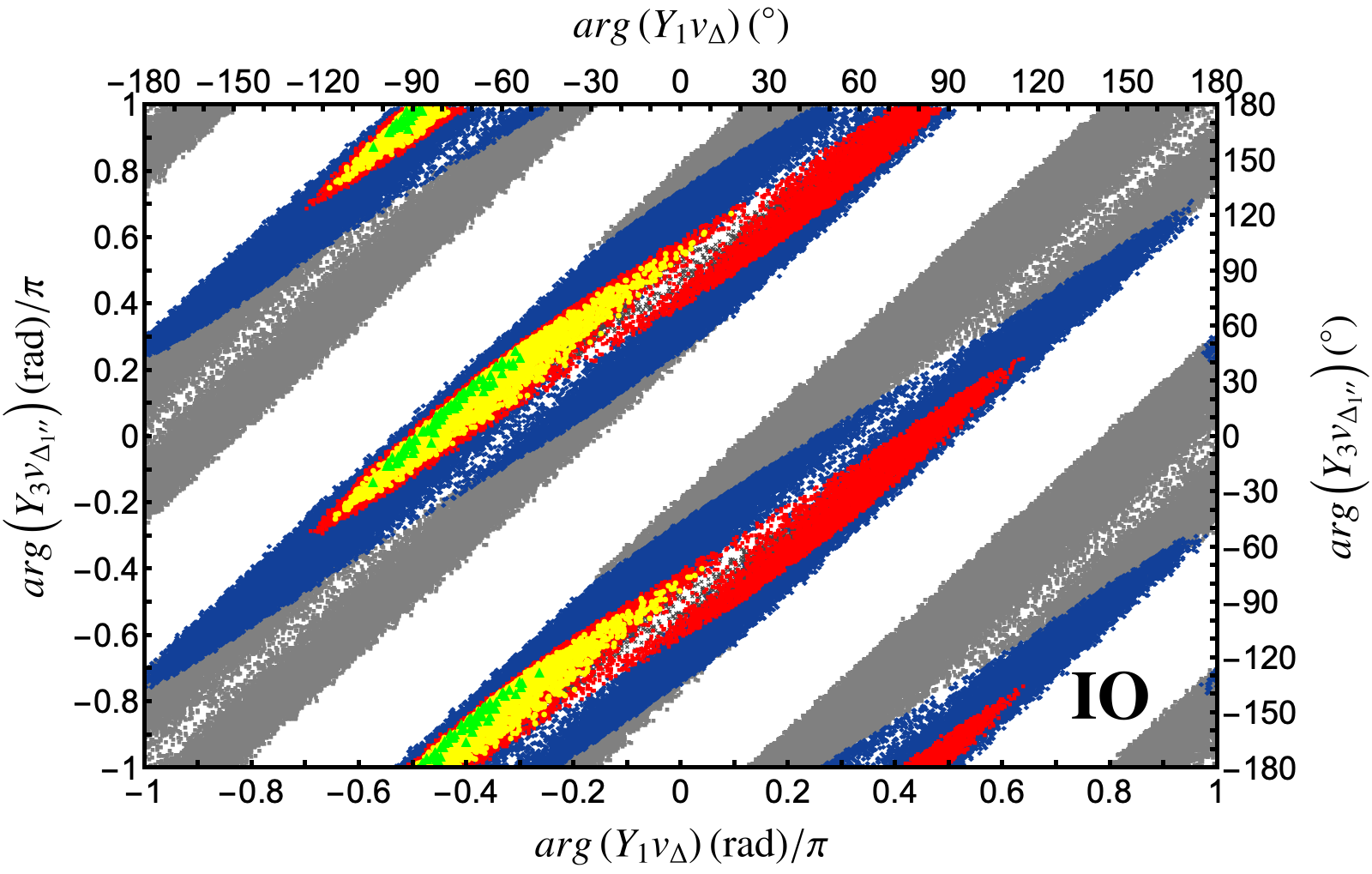}
        \caption{$arg\left(Y_1 v_{\Delta}\right)$ vs $arg\left(Y_3 v_{\Delta_{1^{\prime\prime}}}\right)$ plot for inverted neutrino mass ordering.}
        \label{fig:argY1_argY3_IO}
    \end{subfigure}
    \caption{$arg\left(Y_1 v_{\Delta}\right)$ vs $arg\left(Y_3 v_{\Delta_{1^{\prime\prime}}}\right)$ correlations for normal and inverted neutrino mass ordering. Color labeling is identical to that of plots in the Fig.~\ref{fig:alpha1_alpha2_plots}.}
    \label{fig:argY1_argY3_plots}
\end{figure*}

%$\left|Y_1 v_{\Delta}\right|$ vs $\left|Y_3 v_{\Delta_{1^{\prime\prime}}}\right|$ plots description, Fig. 10
The correlation between Yukawa couplings (See Eq.~\ref{eq:lag_yuk} and Eq.~\ref{eq:yuk_abfgj_relation}) has been analyzed and some results have been shown in Figs.~\ref{fig:absY1_absY3_plots} and~\ref{fig:argY1_argY3_plots}. The former figure shows the plots for $\left|Y_1 v_{\Delta}\right|$ vs $\left|Y_3 v_{\Delta_{1^{\prime\prime}}}\right|$ relation for both normal(See Fig.~\ref{fig:absY1_absY3_NO}) and inverted(See Fig.~\ref{fig:absY1_absY3_IO}) neutrino mass ordering. The obtained results agree with the canonical seesaw-II numbers. The relation between other combinations of absolute values of Yuakwas are similar to the results shown and hence are not shown here. The color legend for a normal and inverted neutrino mass ordering is consistent with other plots (See Fig.~\ref{fig:alpha1_alpha2_plots}).

%$arg\left(Y_1 v_{\Delta}\right)$ vs $arg\left(Y_3 v_{\Delta_{1^{\prime\prime}}}\right)$ plots description, Fig. 11
The later one, See Fig.~\ref{fig:argY1_argY3_plots}, shows the plots for $arg\left(Y_1 v_{\Delta}\right)$ vs $arg\left(Y_3 v_{\Delta_{1^{\prime\prime}}}\right)$ relation for both normal(See Fig.~\ref{fig:argY1_argY3_NO}) and inverted(See Fig.~\ref{fig:argY1_argY3_IO}) neutrino mass ordering. The results of numerical calculations agree with the canonical seesaw-II numbers. The relation between other combinations of phase values of Yuakwas are similar to the results presented and hence are not shown here. The color legend for a normal and inverted neutrino mass ordering is consistent with other plots (See Fig.~\ref{fig:alpha1_alpha2_plots}).

The procedure for generating the correlations between absolute values and phases of Yukawa couplings is as follows: the experimental observables were calculated for the $5\sigma$  ranges of mean values of lepton mixing angles and Dirac phase and for a full range of Majorana phases and the lightest neutrino mass, then only combinations that satisfy the neutrino mass sum rule were taken, then all other experimental constraints described in Sec.~\ref{sec:constraints} were applied, the results that satisfied all the constraints were used to calculate the Yukawa and vacuum expectation values multiplication combinations.

Similar analysis can be performed for $Y_5 v_{s_1}$ vs $Y_6 v_{s_2}$ correlation, since both $Y_5 v_{s_1}$ and $Y_6 v_{s_2}$ are parameteric functions of $Y_4 v_u$. The $Y_4 v_u$ combination is independent parameter in the neutrino flavor mixing analysis. But it can be constrained by the overall neutrino mass scale.
\section{Conclusions}
\label{sec:conclusion}
It has been demonstrated in the present manuscript that $m_{\nu}^{13} + m_{\nu}^{22} + m_{\nu}^{31} = 0$ neutrino mass selection rule can be achieved by a means only one beyond the standard model flavor discrete symmetry, namely the $T^{\prime}$. The required beyond the standard model fields are minimal: $N$ three neutral fermions to accommodate type-I seesaw mechanism, $\Delta$ for type-II seesaw mechanism. Three copies of $\Delta$ and combination of type-I and -II seesaw mechanisms are necessary ingredients for the $m_{\nu}^{13} + m_{\nu}^{22} + m_{\nu}^{31} = 0$ constraints. In case one tries to achieve similar result by means of type-I or type-II seesaw alone, it will require much more fields and symmetries beyond the standard model. $H_{u,d}$ scalar fields contribute to seesaw-I mechanism, whereas $H$ is the canonical standard model Higgs field which, in the present study, gives masses to other standard model particles, except neutrinos. $S_{1,2}$, standard model electroweak singlet scalar fields, are needed to generated the heavy Majorana masses for $N$ fermions.

We highlight some important predictions for the current model that were obtained in the current study: $\alpha_2 \approx 2\pi$ (best fit) for normal neutrino mass ordering; $110^\circ < \alpha_1 < 170^\circ$ and $40(220)^\circ < \alpha_2 < 60(240)^\circ$ for inverted neutrino mass ordering; $240^\circ < \delta_{\text{Dirac}} < 310^\circ$ for the inverted neutrino mass ordering; $m_{\text{lightest}} \approx 3 - 4 \times 10^{-4}$~eV and $m_{ee}^{0\nu\beta\beta} \approx 3 \times 10^{-3}$~eV for normal neutrino mass ordering; $m_{\text{lightest}} \approx 2 \times 10^{-5} - 7 \times 10^{-3}$~eV for inverted neutrino mass ordering.

Future experimental data and improved sensitivities on neutrinoless double beta decay ($m_{ee}^{0\nu\beta\beta}$), $m_{\nu_e}^{\text{eff}}$, and cosmological neutrino mass ($\sum m_{\nu}$) measurements will help further constrain Majorana phases, Dirac phase, lepton mixing angles, the lightest neutrino mass, and determine the neutrino mass ordering scheme for the present $T^{\prime}$ discrete flavor symmetry based model with $m_{\nu}^{13} + m_{\nu}^{22} + m_{\nu}^{31} = 0$ neutrino mass selection rule.
%

%Plots_α1_vs_α2_NO(IO)-----------------------------------------------------------------
\onecolumngrid
% \begin{widetext}
% \end{widetext}
\twocolumngrid

%Plots_α(1)2_vs_δDirac_IO--------------------------------------------------------------
\onecolumngrid
% \twocolumngrid

%Plots_θ12_vs_δDirac_NO--------------------------------------------------------------
%Plots_θ12_vs_θ13_NO-----------------------------------------------------------------
% \onecolumngrid
% \twocolumngrid

%Plots_θ13_vs_δDirac_NO-----------------------------------------------------------------
%Plots_θ23_vs_δDirac_NO-----------------------------------------------------------------
% \onecolumngrid
%\twocolumngrid

%Plots_mν1_vs_0νββ_NO-------------------------------------------------------------------
%Plots_mν3_vs_0νββ_IO-----------------------------------------------------------------
%\onecolumngrid
%\twocolumngrid

%Plots_mν3_vs_KATRIN_NO-------------------------------------------------------------------
%Plots_mν1_vs_KATRIN_IO-----------------------------------------------------------------
%\onecolumngrid
%\twocolumngrid

%Plots_KATRIN_vs_0νββ_NO-------------------------------------------------------------------
%Plots_KATRIN_vs_0νββ_IO-----------------------------------------------------------------
%\onecolumngrid
% \twocolumngrid

%Plots_Σmν_vs_0νββ_NO-------------------------------------------------------------------
%Plots_Σmν_vs_0νββ_IO-----------------------------------------------------------------
%\onecolumngrid
 % \twocolumngrid

%Plots_Σmν_vs_KATRIN_NO-------------------------------------------------------------------
%Plots_Σmν_vs_KATRIN_IO-----------------------------------------------------------------
%\onecolumngrid
% \twocolumngrid

%Plot_absY1_vs_absY3_NO-------------------------------------------------------------------------
%Plot_absY1_vs_absY3_IO-------------------------------------------------------------------------
%\onecolumngrid
% \twocolumngrid

%Plot_argY1_vs_argY3_NO-------------------------------------------------------------------------
%Plot_argY1_vs_argY3_IO-------------------------------------------------------------------------
%\onecolumngrid
\twocolumngrid

%
%-------------------------------------------------------------------------------------------------

\acknowledgments
The work was supported by National Natural Science Fund of China Grant No.~12350410373 (O.~P.) and by the Fundamental Research Funds for Central Universities (T.~N.).
% All Feynman diagrams were created using {\footnotesize TikZ-Feynman LateX} package~\cite{Ellis:2016jkw}.

The order of the authors' names is alphabetical.
%
%--------------------------------------------------------------------------------------------------------------------------------------------------------------------------------------------------------
%
%\appendix
%
%
%\newpage
%\nocite{*}
%
%
%\bibliographystyle{utphys}
\bibliography{references}
\end{document}